\documentclass[prd,preprint,superscriptaddress,showpacs,byrevtex]{revtex4}
\usepackage{bm}
\def\fsl#1{\setbox0=\hbox{$#1$}           
   \dimen0=\wd0                                 
   \setbox1=\hbox{/} \dimen1=\wd1               
   \ifdim\dimen0>\dimen1                        
      \rlap{\hbox to \dimen0{\hfil/\hfil}}      
      #1                                        
   \else                                        
      \rlap{\hbox to \dimen1{\hfil$#1$\hfil}}   
      /                                         
   \fi}                                         %
\newcommand{\be}{\begin{equation}}
\newcommand{\ee}{\end{equation}}
\newcommand{\bea}{\begin{eqnarray}}
\newcommand{\eea}{\end{eqnarray}}
\newcommand{\beq}{\begin{equation}}
\newcommand{\eeq}{\end{equation}}
\newcommand{\beqs}{\begin{eqnarray}}
\newcommand{\eeqs}{\end{eqnarray}}

\begin{document}
\title{ Violation of Factorization Theorem in QCD and QED For Non-Light-Like Wilson Line }
\author{Gouranga C Nayak } \email{nayak@max2.physics.sunysb.edu}
\affiliation{ 458 East College Avenue, \#603, State College, Pennsylvania 16801, USA }
\date{\today}
\begin{abstract}
By using path integral formulation of QCD and QED we prove that the factorization theorem is valid
for light-like Wilson line but is not valid for non-light-like Wilson line. This conclusion is shown
to be consistent with Ward identity and Grammer-Yennie approximation. Hence we find that
the factorization theorem is violated in all the previous studies which used the non-light-like
Wilson line in the definition of  the (transverse momentum dependent)
parton distribution function and fragmentation function at high energy colliders.
\end{abstract}
\maketitle
\pagestyle{plain}
\pagenumbering{arabic}

\section{Introduction}

The Wilson line in QCD is extensively used at high energy colliders. In particular, the
Wilson line is used in the definition of the non-perturbative quantities in QCD such as
in the definition of the parton distribution function and fragmentation function at high
energy colliders \cite{collins}. There are two types of Wilson lines, 1) light-like
Wilson line and 2) non-light-like Wilson line.

In case of light-like Wilson line the parton travels at speed of light and hence
it produces pure gauge potential at all the time-space points $x^\mu$ except at the position
${\vec x}$ transverse to the motion of the parton at the time of closest approach \cite{stermanx,nayakj,nayake}.
Since light-like Wilson line produces pure gauge potential, it produces zero fields {\it i. e.}, $F_{\mu \nu}^a(x)=0$.

In case of non-light-like Wilson line the parton does not travel at speed of light and hence
it does not produce pure gauge potential \cite{stermanx,nayakj,nayake}. Since non-light-like
Wilson line does not produce
pure gauge potential, it produces non-zero fields {\it i. e.}, $F_{\mu \nu}^a(x)\neq 0$.
In addition to non-zero $F_{\mu \nu}^a(x)$, the form of $F_{\mu \nu}^a(x)$ can be arbitrary
for non-light-like Wilson line. Hence one expects fundamental physics differences between
light-like Wilson line and non-light-like Wilson line.

One of such physics issue we will study in this paper is the factorization theorem in QCD
and QED. We will show in this paper, by using path integral formulation of QCD and QED, that the factorization theorem is valid
for light-like Wilson line but is not valid for non-light-like Wilson line.
This conclusion is also consistent with Ward identity and Grammer-Yennie approximation (see section III
for details).

In case of Grammer-Yennie approximation the polarization sums for both real and
virtual photons are rearranged into two parts. One of these is called the $K$-polarization sum
which is proportional to $k^\mu k^\nu$ which corresponds to longitudinal polarization of the
photon field. This corresponds to pure gauge because the polarization of the pure gauge is proportional to its
four momentum $k^\mu$, {\it i. e.}, it is longitudinally polarized (which is unphysical) which
can be gauged away in the sense of factorization \cite{tucci}.

In case of Ward identity the gauge transformation part of the photon field is nothing but the
pure gauge. Since the polarization of the pure gauge is longitudinally polarized (which is unphysical)
it can be gauged away in the sense of factorization (gauge transformation).

Since the light-like Wilson line produces pure gauge and the non-light-like Wilson line does not produce
pure gauge, one finds that the Ward identity and Grammer-Yennie approximation is consistent
with the fact that the factorization theorem is valid
for light-like Wilson line but is not valid for non-light-like Wilson line.

Note that for non-light-like Wilson line
there is no collinear divergence in the Eikonal approximation [see eq. (\ref{scd}) below].
Hence for non-light-like Wilson line one does not have to worry about collinear divergences.
In case of non-light-like Wilson line there can be soft divergences which are not factorized
which we will prove in this paper.

Note that, in some of the studies, the non-light-like Wilson line is used in the definition of the transverse
momentum dependent parton distribution function (TMD PDF) at high energy colliders \cite{collins,collinstmd}.
These studies involve diagrammatic calculation at one loop level using perturbative QCD.
However, the parton distribution function and fragmentation function
are non-perturbative quantities in QCD. Hence if one uses the perturbation theory to study
their properties \cite{collins,collinstmd}, then one may end up finding wrong results.
In general, the non-perturbative phenomena may be impossible to understand by
perturbation theory, regardless of how many orders of perturbation theory one uses. Take for example,
the non-perturbative function
\bea
f(x) =e^{-\frac{1}{x^{2n}}},~~~~~~~~~~n={\rm positive~integer}.
\eea
The Taylor series at $x = 0$ for this function $f(x)$ is exactly zero to all orders in perturbation theory,
but the function is non-zero if $x \neq 0$.

Hence the diagrammatic method using perturbative QCD may not be always sufficient to prove factorization of
soft-collinear divergences of parton distribution function and fragmentation function
at high energy colliders,  regardless of how many orders of perturbation theory one uses,
because parton distribution function and fragmentation function are non-perturbative quantities in QCD.

On the other hand the path integral method of QCD can be used to study non-perturbative QCD.
The path integral method of QCD is valid at all order in coupling constant. Hence we find
that the path integral method of QCD is very useful to prove factorization of
soft-collinear divergences of non-perturbative quantities in QCD such as the
parton distribution function and fragmentation function at all order in coupling constant
at high energy colliders.

In this paper we prove, by using path integral method of QCD, that the factorization theorem in QCD and QED is valid
for light-like Wilson line but is not valid for non-light-like Wilson line.
This implies that the factorization theorem is violated in all the previous studies \cite{collins,collinstmd}
which used the non-light-like Wilson line in the definition of  the (transverse momentum dependent)
parton distribution function and fragmentation function at high energy colliders.
This is in conformation with the finding in \cite{nayaksterman,nayaksterman1} which proved factorization
theorem in NRQCD heavy quarkonium production in case of light-like Wilson line
\cite{nayaksterman} and the violation of factorization theorem in NRQCD heavy quarkonium production in case of
non-light-like Wilson line \cite{nayaksterman1}. In case of massive Wilson line in QCD the color transfer occurs
and the factorization breaks down \cite{nayaksterman1}. The simple physics reason behind this is the following.
The light-like Wilson line produces pure gauge potential which gives $F_{\mu \nu}^a(x) = 0$ which can be gauged away
in the sense of factorization because pure gauge corresponds to unphysical longitudinal polarization
and hence factorization theorem works (see sections IV and VII for details).
The non-light-like Wilson line does not produce pure gauge potential ($F_{\mu \nu}^a(x) \neq 0$),
the form of which can be arbitrary which can not be gauged away
in the sense of factorization because non-pure gauge does not correspond to unphysical longitudinal
polarization and hence the non-light-like Wilson line spoils the factorization (see sections V and VIII for details).

The paper is organized as follows. In section II we briefly discuss soft-collinear divergence and light-like
Wilson line in QED. In section III we discuss Ward Identity, Grammer-Yennie approximation, longitudinal polarization
and light-like Wilson line. In section IV we discuss the proof of factorization of soft-collinear
divergence in QED by using light-like Wilson line. In section V we prove that the soft-collinear
divergence in QED is not factorized when non-light-like Wilson line is used. In section VI we discuss
soft-collinear divergence and light-like Wilson line in QCD. In section VII we discuss
the proof of factorization of soft-collinear divergence in QCD by using light-like Wilson line.
In section VIII we prove that the soft-collinear divergence in QCD is not factorized when non-light-like
Wilson line is used. Section IX contains conclusions.

\section{ Soft-Collinear Divergences and Light-Like Wilson Line in QED }

As mentioned earlier in order to study factorization of soft-collinear divergences the
Wilson line is used in the definition of parton distribution function and fragmentation function
which also make them gauge invariant. Before proceeding to the issue of gauge
invariance and the Wilson line for soft-collinear divergences in QCD let us first discuss
the corresponding situation in QED. The gauge transformation of the Dirac field of the electron in QED is given by
\bea
\psi'(x)=e^{ie\omega(x)}\psi(x).
\label{phg3q}
\eea
Hence we can expect to address the issue of gauge invariance and factorization of soft-collinear divergences in QED
simultaneously if we can relate the $\omega(x)$ to the photon field $A^\mu(x)$. This is easily done by using
Eikonal Feynman rules in QED for soft-collinear divergences when light-like Wilson line is used.
This can be seen as follows.

The Eikonal propagator times the Eikonal vertex for a photon with four momentum $k^\mu$ interacting with an
electron moving with four momentum $p^\mu$ in the limit $k<<< p$ is given by
\cite{collins,tucci,sterman,stermanx,berger,frederix,nayakqed,scet1,pathorder,nayaka2,nayaka3}
\bea
e~\frac{p^\mu}{p \cdot k+i\epsilon }.
\label{eikonaliqp}
\eea
In QED the soft divergence arises only from the
emission of a photon for which all components of the four-momentum $k^\mu$ are small
($ k \rightarrow 0$) which is evident from eq. (\ref{eikonaliqp}). From eq. (\ref{eikonaliqp})
one also finds that when $0<k<<< p$ and ${\vec p}$ is parallel to ${\vec k}$ one may
find collinear divergence.

However, since
\bea
p \cdot k =p_0 k_0 -{\vec p} \cdot {\vec k}=|{\vec p}||{\vec k}|(\sqrt{1+\frac{m^2}{{\vec p}^2}}  -{\rm cos}\theta)
\label{scd}
\eea
one finds that the collinear divergence does not appear in QED because of the non-vanishing mass of the electron,
{\it i. e.}, $m\neq 0$. From eqs. (\ref{eikonaliqp}) and (\ref{scd}) one finds that the collinear divergence appears only
when $m=0$ and $\theta =0$ where $\theta$ is the angle between ${\vec p}$ and ${\vec k}$.
Since gluons are massless and the massless gluons interact with each other one finds that the collinear
divergence appears in QCD. Since a massless particle is always light-like one
finds that the soft-collinear divergences can be described by light-like Wilson line in QCD.

For light-like electron we find from eq. (\ref{eikonaliqp})
\bea
e~\frac{p^\mu}{p \cdot k+i\epsilon }=e~\frac{l^\mu}{l \cdot k+i\epsilon }
\label{eikonaliq}
\eea
where $l^\mu$ is the light-like four-velocity ($|{\vec l}|$=1) of the electron.
Note that when we say the "light-like electron" we mean the electron
that is traveling at its highest speed which is arbitrarily close to the speed of light
($|{\vec l}|\sim 1$) as it can not travel exactly at speed of light
because it has finite mass even if the mass of the electron is very small. Hence we find that
if $l^\mu$ is light-like four velocity then the soft-collinear divergence can be described by
the Eikonal Feynman rule as given by eq. (\ref{eikonaliq}).

From eq. (\ref{eikonaliq}) we find
\bea
&& e\int \frac{d^4k}{(2\pi)^4} \frac{l\cdot {  A}(k)}{l\cdot k +i\epsilon } =-e i\int_0^{\infty} d\lambda \int \frac{d^4k}{(2\pi)^4} e^{i l \cdot k \lambda} l\cdot {A}(k) = ie\int_0^{\infty} d\lambda l\cdot { A}(l\lambda)
\label{ftgtq}
\eea
where the photon field  $ { A}^{\mu }(x)$ and its Fourier transform $ { A}^{\mu }(k)$ are related by
\bea
{ A}^{\mu }(x) =\int \frac{d^4k}{(2\pi)^4} { A}^{\mu }(k) e^{ik \cdot x}.
\label{ftq}
\eea
Now consider the corresponding Feynman diagram for the soft-collinear divergences in QED
due to exchange of two soft-collinear photons of four-momenta $k^\mu_1$ and $k^\mu_2$.
The corresponding Eikonal contribution due to two soft-collinear photons exchange is analogously given by
\bea
&& e^2\int \frac{d^4k_1}{(2\pi)^4} \frac{d^4k_2}{(2\pi)^4} \frac{ l\cdot { A}(k_2) l\cdot { A}(k_1)}{(l\cdot (k_1+k_2) +i \epsilon)(l\cdot k_1 +i \epsilon)} \nonumber \\
&&=e^2i^2 \int_0^{\infty} d\lambda_2 \int_{\lambda_2}^{\infty} d\lambda_1 l\cdot { A}(l\lambda_2) l\cdot {\cal A}(l\lambda_1)
\nonumber \\
&&= \frac{e^2i^2}{2!}\int_0^{\infty} d\lambda_2 \int_0^{\infty} d\lambda_1 l\cdot {\cal A}(l\lambda_2) l\cdot {\cal A}(l\lambda_1).
\eea
Extending this calculation up to infinite number of soft-collinear photons
we find that the Eikonal contribution for the soft-collinear divergences due to
soft-collinear photons exchange with the light-like electron in QED is given by the exponential
\bea
e^{ie \int_0^{\infty} d\lambda l\cdot { A}(l\lambda) }
\label{iiijqed}
\eea
where $l^\mu$ is the light-like four velocity of the electron. The Wilson line in QED is given by
\bea
e^{ie \int_{x_i}^{x_f} dx^\mu A_\mu(x) }.
\label{klj}
\eea
When $A^\mu(x)=A^\mu(l\lambda)$ as in eq. (\ref{iiijqed}) then
one finds from eq. (\ref{klj}) that the light-like Wilson line in QED for soft-collinear divergences is given by \cite{stermanpath}
\bea
e^{ie \int_{x_i}^{x_f} dx^\mu A_\mu(x) }=e^{-ie \int_0^{\infty} d\lambda l\cdot { A}(x_f+l\lambda) }e^{ie \int_0^{\infty} d\lambda l\cdot { A}(x_i+l\lambda) }.
\label{tto}
\eea
Note that a light-like electron traveling with light-like four-velocity $l^\mu$ produces U(1) pure gauge potential $A^{\mu }(x)$
at all the time-space position $x^\mu$ except at the position ${\vec x}$ perpendicular to the direction of motion
of the electron (${\vec l}\cdot {\vec x}=0$) at the time of closest approach \cite{stermanx,nayakj,nayake}.
When $A^{\mu }(x) = A^{\mu }(\lambda l)$ as in eq. (\ref{iiijqed})
we find ${\vec l}\cdot {\vec x}=\lambda {\vec l}\cdot {\vec l}=\lambda\neq 0$ which implies that the light-like Wilson line
finds the photon field $A^{\mu }(x)$ in eq. (\ref{iiijqed}) as the U(1) pure gauge. The U(1) pure gauge is given by
\bea
A^\mu(x)=\partial^\mu \omega(x)
\label{purea1}
\eea
which gives from eq. (\ref{tto}) the light-like Wilson line in QED for soft-collinear divergences
\bea
e^{ie\omega(x_f)}e^{-ie\omega(x_i)}=e^{ie \int_{x_i}^{x_f} dx^\mu A_\mu(x) }=e^{-ie \int_0^{\infty} d\lambda l\cdot { A}(x_f+l\lambda) }e^{ie \int_0^{\infty} d\lambda l\cdot { A}(x_i+l\lambda) }
\label{lkj}
\eea
which depends only on end points $x_i^\mu$ and $x_f^\mu$ but is independent of the path. The path independence can also be found from
Stokes theorem because for pure gauge
\bea
F^{\mu \nu}(x)=\partial^\mu A^\nu(x)-\partial^\nu A^\mu(x)=0
\eea
which gives from Stokes theorem
\bea
e^{ie \oint_C dx^\mu A_\mu(x) }=e^{ie \int_S dx^\mu dx^\nu F_{\mu \nu}(x) }=1
\eea
where $C$ is a closed path and $S$ is the surface enclosing $C$. Now considering two different paths
$L$ and $M$ with common end points $x_i^\mu$ and $x_f^\mu$ we find
\bea
e^{ie \oint_C dx^\mu A_\mu(x) }=e^{ie \int_L dx^\mu A_\mu(x)- ie \int_M dx^\mu A_\mu(x)}=1
\eea
which implies that
\bea
e^{ie \int_{x_i}^{x_f} dx^\mu A_\mu(x) }
\eea
depends only on end points $x_i^\mu$ and $x_f^\mu$ but is independent of path which can also be seen from eq. (\ref{lkj}).
Hence from eq. (\ref{lkj}) we find that the abelian phase or the gauge link in QED is given by
\bea
e^{-ie \int_0^{\infty} d\lambda l\cdot { A}(x+l\lambda) }=e^{ie\omega(x)}.
\label{phas}
\eea
Eq. (\ref{phas}) can also be directly verified as follows. When $A^\mu(x)=\partial^\mu \omega(x)$ is U(1)
pure gauge as given by eq. (\ref{purea1}) we find by using contour integration that
\bea
&&e^{-ie \int_0^{\infty} d\lambda l\cdot  A(x+l\lambda) }=e^{-ie \int_0^{\infty} d\lambda  e^{\lambda l \cdot \partial}~l \cdot  A(x) } =e^{-ie \int_0^{\infty} d\lambda  e^{\lambda l \cdot \partial}~\int \frac{d^4k}{(2\pi)^4} e^{ik \cdot x} l \cdot  A(k) } \nonumber \\
&& =e^{-ie \int \frac{d^4k}{(2\pi)^4} \int_0^{\infty} d\lambda  e^{i\lambda l \cdot k}~e^{ik \cdot x} l \cdot  A(k) }=e^{ie \int \frac{d^4k}{(2\pi)^4}  \frac{1}{il \cdot k }~e^{ik \cdot x} l \cdot  A(k) }=e^{ie \frac{1}{l \cdot \partial }~ l \cdot  A(x) }=e^{ie\omega(x) }
\label{ppgond}
\eea
which reproduces eq. (\ref{phas}).

From eqs. (\ref{phg3q}) and (\ref{phas}) one expects that the gauge invariance and factorization of
soft-collinear divergences in QED can be explained simultaneously if the light-like Wilson line is
used.

\section{ Ward Identity, Grammer-Yennie approximation, longitudinal polarization and light-like Wilson line}

Note that, as mentioned above, in classical electrodynamics the assertion that the gauge field that is produced
by a highly relativistic particle is a pure gauge \cite{stermanx,nayakj,nayake}.
In order to study factorization of infrared divergences by using the background field method of QED,
the soft photon cloud traversed by the electron
is represented by the background field $A^\mu(x)$ \cite{tucci} where one represents the (hard) quantum
photon field by $Q^\mu(x)$.

In order to see how, in the quantum field theory (QED),
the pure gauge property holds for the case in which the photon momentum is soft
can be seen from the Grammer-Yennie approximation \cite{grammer},
see also \cite{tucci}. In the Grammer-Yennie approximation, if the photon is emitted from an incoming line,
then for soft-divergence one replaces the polarization of the gauge field
[see eq. (3.2) of \cite{grammer}] by
\begin{equation}
\epsilon^\mu(k) = [\epsilon^\mu(k) -k^\mu \frac{l \cdot \epsilon(k)}{l \cdot k}]+k^\mu \frac{l \cdot \epsilon(k)}{l \cdot k}
\label{1}
\end{equation}
which, when used in the Feynman rule for the eikonal line [that is given in eq. (\ref{eikonaliq}) above or in
the equation before eq. (3.2) of \cite{grammer}], gives
\begin{equation}
\frac{l \cdot \epsilon(k)}{l \cdot k} = [ \frac{l \cdot \epsilon(k)}{l \cdot k}- \frac{l \cdot \epsilon(k)}{l \cdot k}]+ \frac{l \cdot \epsilon(k)}{l \cdot k} =0+\frac{l \cdot \epsilon(k)}{l \cdot k}.
\label{2}
\end{equation}
Hence one finds that the first term $[\epsilon^\mu(k) -k^\mu \frac{l \cdot \epsilon(k)}{l \cdot k}]$ in the right hand side
of eq. (\ref{1}) can not lead to soft divergence. The soft divergence comes
from the last term $k^\mu \frac{l \cdot \epsilon(k)}{l \cdot k}$ in the right hand side of eq. (\ref{1}).
Hence from the Grammer-Yennie approximation one finds that, as far as soft divergence is concerned, the polarization of the gauge field
is proportional to $k^\mu$, {\it i. e.}, it is longitudinally polarized.
Since the polarization of the pure gauge is proportional to $k^\mu$,
one finds that the pure gauge property holds for the case in which the gauge field momentum
is soft. Since longitudinal polarizations are unphysical they can be gauged away, in the sense
of factorization \cite{tucci}. Hence one finds that the factorization theorem for soft/infrared divergences can be studied by using pure gauge.

The pure gauge property is also evident from the Ward-Identity which can be seen when one replaces
$\epsilon^\mu(k)$ in the QED amplitude
\bea
{\cal M} = \epsilon^\mu(k) {\cal M}_\mu(k)
\eea
by
\bea
\epsilon^\mu(k) \rightarrow k^\mu~~~~~~~~~~~~{\rm Pure~gauge}
\eea
then one gets
\bea
k^\mu {\cal M}_\mu(k) =0
\eea
which eliminates longitudinal polarization of the photon in the physical process. This is the statement of gauge transformation
which is equivalent to saying that the gauge transformation part is noting but the pure gauge. Hence the statement that unphysical
longitudinal polarization can be gauged away by using pure gauge, in the sense of factorization \cite{tucci}, is consistent with Ward-identity.

The above analysis in the quantum field theory is also valid for collinear divergence in the Eikonal approximation
for light-like Wilson line case because in the collinear limit the momentum $k^\mu$ of the gauge field is proportional to the
four velocity of the light-like Wilson line $l^\mu$. Since the polarization of the pure gauge is proportional to $k^\mu$,
one finds that the pure gauge property holds for the case in which the gauge field momentum
is collinear to the momentum of the light-like Wilson line.

Note that for non-light-like Wilson line
there is no collinear divergence in the Eikonal approximation which can be seen from eq. (\ref{scd}).
Hence for non-light-like Wilson line one does not have to worry about collinear divergences.

In case of non-light-like Wilson line there can be soft divergences which are not factorized
which we will prove in this paper. This is consistent with the fact that non-light-like Wilson
line can not produce pure gauge \cite{stermanx,nayakj,nayake}.

Hence, since the light-like Wilson line produces pure gauge and the non-light-like Wilson line does not produce
pure gauge \cite{stermanx,nayakj,nayake}, one finds that the Ward identity and Grammer-Yennie approximation is consistent
with the fact that the factorization theorem is valid
for light-like Wilson line but is not valid for non-light-like Wilson line.

\section{Proof of Factorization of Soft-Collinear Divergences in QED With Light-Like Wilson Line }

Note that the gauge invariant greens function in QED
\bea
G(x_1,x_2)=<{\bar \psi}(x_2)~\times~{\rm exp}[ie \int_{x_1}^{x_2} dx^\mu A_\mu(x)]~\times~ \psi(x_1)>
\label{qed}
\eea
in the presence of background field $A^\mu(x)$ was formulated by Schwinger long time ago \cite{schw1}.
When this background field $A^\mu(x)$ is replaced by the U(1) pure gauge background field
as given by eq. (\ref{purea1}) [which corresponds to the light-like Wilson line in eq. (\ref{phas})]
then the path integral method of QED can be used to prove
gauge invariance and factorization of soft-collinear divergences in QED simultaneously.

\subsection{ Proof of Factorization Theorem in QED With Light-Like Wilson Line in Covariant Gauge }

The generating functional in QED in the path integral formulation in covariant gauge is given by
\bea
Z[J,\eta,{\bar \eta}]=\int [dQ] [d{\bar \psi}] [d \psi ]~
e^{i\int d^4x [-\frac{1}{4}{F}_{\mu \nu}^2[Q] -\frac{1}{2 \alpha} (\partial_\mu Q^{\mu })^2+{\bar \psi} [i\gamma^\mu \partial_\mu -m +e\gamma^\mu Q_\mu] \psi+ J \cdot Q +{\bar \eta} \psi + {\bar \psi} \eta ]}
\label{zfqv}
\eea
where $J^\mu(x)$ is the external source for the quantum photon field $Q^\mu(x)$ and ${\bar \eta}(x)$ is the external source for the
Dirac field $\psi(x)$ of the electron and
\bea
F^{\mu \nu}[Q]=\partial^\mu Q^\nu(x)-\partial^\nu Q^\mu(x),~~~~~~~~~{F}_{\mu \nu}^2[Q]={F}^{\mu \nu}[Q]{F}_{\mu \nu}[Q].
\eea
The correlation function of the type $<0|{\bar \psi}(x_2)\psi(x_1)|0>$ in QED in covariant gauge is given by \cite{tucci}
\bea
&&<0|{\bar \psi}(x_2)\psi(x_1)|0>=\int [dQ] [d{\bar \psi}] [d \psi ]~{\bar \psi}(x_2)\psi(x_1) e^{i\int d^4x [-\frac{1}{4}{F}_{\mu \nu}^2[Q] -\frac{1}{2 \alpha} (\partial_\mu Q^{\mu })^2+{\bar \psi} [i\gamma^\mu \partial_\mu -m +e\gamma^\mu Q_\mu]\psi]}\nonumber \\
\label{zfqvc}
\eea
where suppression of the factor $\frac{1}{Z[0]}$ is understood as it will cancel in the final result (see eq. (\ref{llfac})).

The generating functional in the background field method of QED in covariant gauge is given by
\bea
Z[A,J,\eta,{\bar \eta}]=\int [dQ] [d{\bar \psi}] [d \psi ]~
e^{i\int d^4x [-\frac{1}{4}{F}_{\mu \nu}^2[Q] -\frac{1}{2 \alpha} (\partial_\mu Q^{\mu })^2+{\bar \psi}  [i\gamma^\mu \partial_\mu -m +e\gamma^\mu (A+Q)_\mu]  \psi+ J \cdot Q +{\bar \eta} \psi + {\bar \psi}\eta  ]}\nonumber \\
\label{zfqi}
\eea
where we have used the notation $Q^\mu(x)$ for the quantum photon field and $A^\mu(x)$ for the background field.
The correlation function of the type $<0|{\bar \psi}(x_2)\psi(x_1)|0>$ in the background field method of QED in
covariant gauge is given by \cite{tucci}
\bea
&&<0|{\bar \psi}(x_2)\psi(x_1)|0>_A=\int [dQ] [d{\bar \psi}] [d \psi ]~{\bar \psi}(x_2)\psi(x_1) e^{i\int d^4x [-\frac{1}{4}{F}_{\mu \nu}^2[Q] -\frac{1}{2 \alpha} (\partial_\mu Q^{\mu })^2+{\bar \psi}  [i\gamma^\mu \partial_\mu -m +e\gamma^\mu (A+Q)_\mu]  \psi]}\nonumber \\
\label{zfqvcb}
\eea
where suppression of the factor $\frac{1}{Z[0]}$ is understood as it will cancel in the final result (see eq. (\ref{llfac})).
We write eq. (\ref{zfqvcb}) as
\bea
<0|{\bar \psi}(x_2)\psi(x_1)|0>_A=\int [dQ] [d{\bar \psi}'] [d \psi' ]~{\bar \psi}'(x_2)\psi'(x_1)
e^{i\int d^4x [-\frac{1}{4}{F}_{\mu \nu}^2[Q] -\frac{1}{2 \alpha} (\partial_\mu Q^{\mu })^2+{\bar \psi}' [i\gamma^\mu \partial_\mu -m +e\gamma^\mu (A+Q)_\mu]  \psi']}\nonumber \\
\label{zfqia}
\eea
because the change of integration variable from unprimed to primed variable does not change the value of the integration.

The light-like Wilson line in the background field method of QED is given by, see eqs. (\ref{tto}) and (\ref{phas})
\bea
e^{ie\int_0^x dx'^\mu A_\mu(x')}
\label{a}
\eea
where $A^\mu(x)$ is the U(1) pure gauge potential as given by eq. (\ref{purea1}). Hence for light-like Wilson line we find
\bea
\int_0^x dx'^\nu A_\nu(x')= \int_0^1 d\tau \frac{d\omega(x'(\tau))}{d\tau}=\omega(x)-\omega(0), ~~~~~~~x'(1)=x,~~~~~~~x'(0)=0
\eea
which gives
\bea
\partial_\mu [\int_0^x dx'^\nu A_\nu(x')] = \partial_\mu \omega(x)=A_\mu(x).
\label{jjj}
\eea
From eq. (\ref{phas}) let us define
\bea
\psi'(x) = e^{-ie \int_0^{\infty} d\lambda l\cdot { A}(x+l\lambda) }\psi(x) =e^{ie\omega(x)} \psi(x).
\label{ppg}
\eea
From eq. (\ref{ppgond}) we find for light-like Wilson line [$A^\mu(x)=\partial^\mu\omega(x)$] that
\bea
 \partial_\mu [\int_0^{\infty} d\lambda l\cdot  A(x+l\lambda)]  = -\partial_\mu [\frac{1}{l \cdot \partial }~ l \cdot  A(x)] =- \partial_\mu \omega(x) =-A_\mu(x).
\label{ppgondc}
\eea

Hence when the  background field $A^\mu(x)$ is the U(1) pure gauge as given by eq. (\ref{purea1}), which is
the case for light-like Wilson line, we find from eqs. (\ref{ppg}), (\ref{ppgondc}) and (\ref{zfqia}) that
\bea
&& <0|{\bar \psi}(x_2)\psi(x_1)|0>_A=\int [dQ] [d{\bar \psi}] [d \psi ]~{\bar \psi}(x_2)e^{ie \int_0^{\infty} d\lambda l\cdot { A}(x_2+l\lambda) }e^{-ie \int_0^{\infty} d\lambda l\cdot { A}(x_1+l\lambda) }\psi(x_1)\nonumber \\
&&\times e^{i\int d^4x [-\frac{1}{4}{F}_{\mu \nu}^2[Q] -\frac{1}{2 \alpha} (\partial_\mu Q^{\mu })^2+{\bar \psi} [i\gamma^\mu \partial_\mu -m +e\gamma^\mu Q_\mu]  \psi ]}
\label{zfqibx}
\eea
which gives by using eq. (\ref{lkj})
\bea
&& <0|{\bar \psi}(x_2)e^{ie \int_{x_1}^{x_2} dx^\mu A_\mu(x)}\psi(x_1)|0>_A=\int [dQ] [d{\bar \psi}] [d \psi ]~{\bar \psi}(x_2)\psi(x_1)\nonumber \\
&&\times e^{i\int d^4x [-\frac{1}{4}{F}_{\mu \nu}^2[Q] -\frac{1}{2 \alpha} (\partial_\mu Q^{\mu })^2+{\bar \psi} [i\gamma^\mu \partial_\mu -m +e\gamma^\mu Q_\mu]  \psi ]}.
\label{zfqibk}
\eea
Hence from eqs. (\ref{zfqibk}) and (\ref{zfqvc}) we find that
\bea
<0|{\bar \psi}(x_2)\psi(x_1)|0>=<0|{\bar \psi}(x_2)e^{ie \int_{x_1}^{x_2} dx^\mu A_\mu(x)}\psi(x_1)|0>_A
\label{llfac}
\eea
where $ <0|{\bar \psi}(x_2)e^{ie \int_{x_1}^{x_2} dx^\mu A_\mu(x)}\psi(x_1)|0>_A$ is gauge invariant.
By using eq. (\ref{lkj}) we find from eq. (\ref{llfac}) that
\bea
<{\bar \psi}(x_2) \psi(x_1)>=<{\bar \psi}(x_2)~e^{-ie\int_0^{\infty} d\lambda l\cdot { A}(x_2+l\lambda)}e^{ie\int_0^{\infty} d\lambda l\cdot { A}(x_1+l\lambda)} \psi(x_1)>_A
\label{tuccix}
\eea
which proves factorization of soft-collinear divergences in QED at all order in coupling constant in covariant gauge
when light-like Wilson line is used.
The eq. (\ref{tuccix}) was first proved by Tucci  in \cite{tucci}. We have included its derivation in this section because we will follow
the similar path integral technique in the next section to prove that the factorization theorem is violated in QED if we replace the
light-like Wilson line by the non-light-like Wilson line.

\subsection{ Proof of Factorization Theorem in QED With Light-Like Wilson Line in General Axial Gauge }

The factorization of soft-collinear divergences in QED at all order in coupling constant in covariant gauge
using light-like Wilson line which we proved in the previous sub-section can also be proved in other gauges.
The generating functional in QED in the path integral formulation in general axial gauge is given by \cite{leib,meis}
\bea
Z[J,\eta,{\bar \eta}]=\int [dQ] [d{\bar \psi}] [d \psi ]~
e^{i\int d^4x [-\frac{1}{4}{F}_{\mu \nu}^2[Q] -\frac{1}{2 \alpha} (\eta_\mu Q^{\mu })^2+{\bar \psi} [i\gamma^\mu \partial_\mu -m +e\gamma^\mu Q_\mu] \psi+ J \cdot Q +{\bar \eta} \psi + {\bar \psi} \eta ]}
\label{zfqvaa}
\eea
where $\eta^\mu$ is an arbitrary but constant four vector
\bea
&&\eta^2 <0,~~~~~~~~{\rm axial~gauge}\nonumber \\
&& \eta^2 =0,~~~~~~~~{\rm light}-{\rm cone~gauge}\nonumber \\
&& \eta^2 >0,~~~~~~~~{\rm temporal~gauge}.
\label{axial}
\eea
The correlation function of the type $<0|{\bar \psi}(x_2)\psi(x_1)|0>$ in QED in general axial gauge is given by \cite{meis}
\bea
&&<0|{\bar \psi}(x_2)\psi(x_1)|0>=\int [dQ] [d{\bar \psi}] [d \psi ]~{\bar \psi}(x_2)\psi(x_1) e^{i\int d^4x [-\frac{1}{4}{F}_{\mu \nu}^2[Q] -\frac{1}{2 \alpha} (\eta_\mu Q^{\mu })^2+{\bar \psi} [i\gamma^\mu \partial_\mu -m +e\gamma^\mu Q_\mu]\psi]}.\nonumber \\
\label{zfqvca}
\eea

The generating functional in the background field method of QED in general axial gauge is given by
\bea
Z[A,J,\eta,{\bar \eta}]=\int [dQ] [d{\bar \psi}] [d \psi ]~
e^{i\int d^4x [-\frac{1}{4}{F}_{\mu \nu}^2[Q] -\frac{1}{2 \alpha} (\eta_\mu Q^{\mu })^2+{\bar \psi}  [i\gamma^\mu \partial_\mu -m +e\gamma^\mu (A+Q)_\mu]  \psi+ J \cdot Q +{\bar \eta} \psi + {\bar \psi}\eta  ]}.\nonumber \\
\label{zfqiaxz}
\eea
The correlation function of the type $<0|{\bar \psi}(x_2)\psi(x_1)|0>$ in the background field method of QED in general axial gauge is given by
\bea
&&<0|{\bar \psi}(x_2)\psi(x_1)|0>_A=\int [dQ] [d{\bar \psi}] [d \psi ]~{\bar \psi}(x_2)\psi(x_1) e^{i\int d^4x [-\frac{1}{4}{F}_{\mu \nu}^2[Q] -\frac{1}{2 \alpha} (\eta_\mu Q^{\mu })^2+{\bar \psi}  [i\gamma^\mu \partial_\mu -m +e\gamma^\mu (A+Q)_\mu]  \psi]}.\nonumber \\
\label{zfqvcba}
\eea
We write eq. (\ref{zfqvcba}) as
\bea
<0|{\bar \psi}(x_2)\psi(x_1)|0>_A=\int [dQ] [d{\bar \psi}'] [d \psi' ]~{\bar \psi}'(x_2)\psi'(x_1)
e^{i\int d^4x [-\frac{1}{4}{F}_{\mu \nu}^2[Q] -\frac{1}{2 \alpha} (\eta_\mu Q^{\mu })^2+{\bar \psi}' [i\gamma^\mu \partial_\mu -m +e\gamma^\mu (A+Q)_\mu]  \psi']}\nonumber \\
\label{zfqiaa}
\eea
because the change of integration variable from unprimed to primed variable does not change the value of the integration.

Hence when the  background field $A^\mu(x)$ is the U(1) pure gauge as given by eq. (\ref{purea1}), which is
the case for light-like Wilson line, we find from eqs. (\ref{ppg}), (\ref{ppgondc}) and (\ref{zfqiaa}) that
\bea
&& <0|{\bar \psi}(x_2)\psi(x_1)|0>_A=\int [dQ] [d{\bar \psi}] [d \psi ]~{\bar \psi}(x_2)e^{ie \int_0^{\infty} d\lambda l\cdot { A}(x_2+l\lambda) }e^{-ie \int_0^{\infty} d\lambda l\cdot { A}(x_1+l\lambda) }\psi(x_1)\nonumber \\
&&\times e^{i\int d^4x [-\frac{1}{4}{F}_{\mu \nu}^2[Q] -\frac{1}{2 \alpha} (\eta_\mu Q^{\mu })^2+{\bar \psi} [i\gamma^\mu \partial_\mu -m +e\gamma^\mu Q_\mu]  \psi ]}
\label{zfqibxa}
\eea
which gives by using eq. (\ref{lkj})
\bea
&& <0|{\bar \psi}(x_2)e^{ie \int_{x_1}^{x_2} dx^\mu A_\mu(x)}\psi(x_1)|0>_A=\int [dQ] [d{\bar \psi}] [d \psi ]~{\bar \psi}(x_2)\psi(x_1)\nonumber \\
&&\times e^{i\int d^4x [-\frac{1}{4}{F}_{\mu \nu}^2[Q] -\frac{1}{2 \alpha} (\eta_\mu Q^{\mu })^2+{\bar \psi} [i\gamma^\mu \partial_\mu -m +e\gamma^\mu Q_\mu]  \psi ]}.
\label{zfqibaa}
\eea
Hence from eqs. (\ref{zfqibaa}) and (\ref{zfqvca}) we find that
\bea
<0|{\bar \psi}(x_2)\psi(x_1)|0>=<0|{\bar \psi}(x_2)e^{ie \int_{x_1}^{x_2} dx^\mu A_\mu(x)}\psi(x_1)|0>_A
\label{llfacak}
\eea
where $ <0|{\bar \psi}(x_2)e^{ie \int_{x_1}^{x_2} dx^\mu A_\mu(x)}\psi(x_1)|0>_A$ is gauge invariant.
By using eq. (\ref{lkj}) we find from eq. (\ref{llfacak}) that
\bea
<{\bar \psi}(x_2) \psi(x_1)>=<{\bar \psi}(x_2)~e^{-ie\int_0^{\infty} d\lambda l\cdot { A}(x_2+l\lambda)}e^{ie\int_0^{\infty} d\lambda l\cdot { A}(x_1+l\lambda)} \psi(x_1)>_A
\label{tuccixa}
\eea
which proves factorization of soft-collinear divergences in QED at all order in coupling constant in general axial
gauge when light-like Wilson line is used.

\subsection{ Proof of Factorization Theorem in QED With Light-Like Wilson Line in Light-Cone Gauge }

The light-cone gauge corresponds to \cite{leib,meis,collinsp}
\bea
\eta \cdot Q=0,~~~~~~~~~~~~~~~~\eta^2=0
\label{jnk}
\eea
which is already covered by eqs. (\ref{zfqvaa}) and (\ref{axial}) where the corresponding gauge fixing term
is given by -$\frac{1}{2 \alpha} (\eta_\mu Q^{\mu })^2$. In the light-cone coordinate system
the light-cone gauge \cite{collinsp}
\bea
Q^{+}=0
\label{lightc}
\eea
corresponds to
\bea
\eta^\mu = (\eta^+,\eta^-,\eta_\perp)= (0,1,0)
\label{ligtc}
\eea
which covers $\eta \cdot Q=0$ and $\eta^2=0$ situation in eq. (\ref{jnk}).

Since eq. (\ref{tuccixa}) is valid for general axial gauges [as given by eq. (\ref{axial})] it is also valid for
the light cone gauge $\eta^2=0$. Hence from eq. (\ref{tuccixa}) we find in the light cone gauge that
\bea
<{\bar \psi}(x_2) \psi(x_1)>=<{\bar \psi}(x_2)~e^{-ie\int_0^{\infty} d\lambda l\cdot { A}(x_2+l\lambda)}e^{ie\int_0^{\infty} d\lambda l\cdot { A}(x_1+l\lambda)} \psi(x_1)>_A
\label{tuccixanlb}
\eea
which proves factorization of soft-collinear divergences in QED at all order in coupling constant in light cone
gauge when light-like Wilson line is used.

\subsection{ Proof of Factorization Theorem in QED With Light-Like Wilson Line in General Non-Covariant Gauge }

The generating functional in QED in the path integral formulation in general non-covariant gauge is given by \cite{noncov,noncov1}
\bea
Z[J,\eta,{\bar \eta}]=\int [dQ] [d{\bar \psi}] [d \psi ]~
e^{i\int d^4x [-\frac{1}{4}{F}_{\mu \nu}^2[Q] -\frac{1}{2 \alpha} (\frac{\eta^\mu \eta^\nu}{\eta^2}\partial_\mu Q_\nu)^2+{\bar \psi} [i\gamma^\mu \partial_\mu -m +e\gamma^\mu Q_\mu] \psi+ J \cdot Q +{\bar \eta} \psi + {\bar \psi} \eta ]}
\label{zfqvaan}
\eea
where $\eta^\mu$ is an arbitrary but constant four vector.
The correlation function of the type $<0|{\bar \psi}(x_2)\psi(x_1)|0>$ in QED in general non-covariant gauge is given by
\bea
&&<0|{\bar \psi}(x_2)\psi(x_1)|0>=\int [dQ] [d{\bar \psi}] [d \psi ]~{\bar \psi}(x_2)\psi(x_1) e^{i\int d^4x [-\frac{1}{4}{F}_{\mu \nu}^2[Q] -\frac{1}{2 \alpha} (\frac{\eta^\mu \eta^\nu}{\eta^2}\partial_\mu Q_\nu)^2+{\bar \psi} [i\gamma^\mu \partial_\mu -m +e\gamma^\mu Q_\mu]\psi]}.\nonumber \\
\label{zfqvcan}
\eea

The generating functional in the background field method of QED in general non-covariant gauge is given by
\bea
Z[A,J,\eta,{\bar \eta}]=\int [dQ] [d{\bar \psi}] [d \psi ]~
e^{i\int d^4x [-\frac{1}{4}{F}_{\mu \nu}^2[Q] -\frac{1}{2 \alpha} (\frac{\eta^\mu \eta^\nu}{\eta^2}\partial_\mu Q_\nu)^2+{\bar \psi}  [i\gamma^\mu \partial_\mu -m +e\gamma^\mu (A+Q)_\mu]  \psi+ J \cdot Q +{\bar \eta} \psi + {\bar \psi}\eta  ]}.\nonumber \\
\label{zfqiaxzn}
\eea
The correlation function of the type $<0|{\bar \psi}(x_2)\psi(x_1)|0>$ in the background field method of QED in general non-covariant gauge
is given by
\bea
&&<0|{\bar \psi}(x_2)\psi(x_1)|0>_A=\int [dQ] [d{\bar \psi}] [d \psi ]~{\bar \psi}(x_2)\psi(x_1) \nonumber \\
&& e^{i\int d^4x [-\frac{1}{4}{F}_{\mu \nu}^2[Q] -\frac{1}{2 \alpha} (\frac{\eta^\mu \eta^\nu}{\eta^2}\partial_\mu Q_\nu)^2+{\bar \psi}  [i\gamma^\mu \partial_\mu -m +e\gamma^\mu (A+Q)_\mu]  \psi]}.
\label{zfqvcban}
\eea
We write eq. (\ref{zfqvcban}) as
\bea
&&<0|{\bar \psi}(x_2)\psi(x_1)|0>_A=\int [dQ] [d{\bar \psi}'] [d \psi' ]~{\bar \psi}'(x_2)\psi'(x_1)\nonumber \\
&&e^{i\int d^4x [-\frac{1}{4}{F}_{\mu \nu}^2[Q] -\frac{1}{2 \alpha} (\frac{\eta^\mu \eta^\nu}{\eta^2}\partial_\mu Q_\nu)^2+{\bar \psi}' [i\gamma^\mu \partial_\mu -m +e\gamma^\mu (A+Q)_\mu]  \psi']}
\label{zfqiaan}
\eea
because the change of integration variable from unprimed to primed variable does not change the value of the integration.

Hence when the  background field $A^\mu(x)$ is the U(1) pure gauge as given by eq. (\ref{purea1}), which is
the case for light-like Wilson line, we find from eqs. (\ref{ppg}), (\ref{ppgondc}) and (\ref{zfqiaan}) that
\bea
&& <0|{\bar \psi}(x_2)\psi(x_1)|0>_A=\int [dQ] [d{\bar \psi}] [d \psi ]~{\bar \psi}(x_2)e^{ie \int_0^{\infty} d\lambda l\cdot { A}(x_2+l\lambda) }e^{-ie \int_0^{\infty} d\lambda l\cdot { A}(x_1+l\lambda) }\psi(x_1)\nonumber \\
&&\times e^{i\int d^4x [-\frac{1}{4}{F}_{\mu \nu}^2[Q] -\frac{1}{2 \alpha} (\frac{\eta^\mu \eta^\nu}{\eta^2}\partial_\mu Q_\nu)^2+{\bar \psi} [i\gamma^\mu \partial_\mu -m +e\gamma^\mu Q_\mu]  \psi ]}
\label{zfqibxan}
\eea
which gives by using eq. (\ref{lkj})
\bea
&& <0|{\bar \psi}(x_2)e^{ie \int_{x_1}^{x_2} dx^\mu A_\mu(x)}\psi(x_1)|0>_A=\int [dQ] [d{\bar \psi}] [d \psi ]~{\bar \psi}(x_2)\psi(x_1)\nonumber \\
&&\times e^{i\int d^4x [-\frac{1}{4}{F}_{\mu \nu}^2[Q] -\frac{1}{2 \alpha} (\frac{\eta^\mu \eta^\nu}{\eta^2}\partial_\mu Q_\nu)^2+{\bar \psi} [i\gamma^\mu \partial_\mu -m +e\gamma^\mu Q_\mu]  \psi ]}.
\label{zfqibaan}
\eea
Hence from eqs. (\ref{zfqibaan}) and (\ref{zfqvcan}) we find that
\bea
<0|{\bar \psi}(x_2)\psi(x_1)|0>=<0|{\bar \psi}(x_2)e^{ie \int_{x_1}^{x_2} dx^\mu A_\mu(x)}\psi(x_1)|0>_A
\label{llfacakn}
\eea
where $ <0|{\bar \psi}(x_2)e^{ie \int_{x_1}^{x_2} dx^\mu A_\mu(x)}\psi(x_1)|0>_A$ is gauge invariant.
By using eq. (\ref{lkj}) we find from eq. (\ref{llfacakn}) that
\bea
<{\bar \psi}(x_2) \psi(x_1)>=<{\bar \psi}(x_2)~e^{-ie\int_0^{\infty} d\lambda l\cdot { A}(x_2+l\lambda)}e^{ie\int_0^{\infty} d\lambda l\cdot { A}(x_1+l\lambda)} \psi(x_1)>_A
\label{tuccixan}
\eea
which proves factorization of soft-collinear divergences in QED at all order in coupling constant in general non-covariant
gauge when light-like Wilson line is used.

\subsection{ Proof of Factorization Theorem in QED With Light-Like Wilson Line in General Coulomb Gauge }

The generating functional in QED in the path integral formulation in general Coulomb gauge is given by \cite{noncov,noncov1}
\bea
Z[J,\eta,{\bar \eta}]=\int [dQ] [d{\bar \psi}] [d \psi ]~
e^{i\int d^4x [-\frac{1}{4}{F}_{\mu \nu}^2[Q] -\frac{1}{2 \alpha} ([g^{\mu \nu}-\frac{n^\mu n^\nu}{n^2}]\partial_\mu Q_\nu)^2+{\bar \psi} [i\gamma^\mu \partial_\mu -m +e\gamma^\mu Q_\mu] \psi+ J \cdot Q +{\bar \eta} \psi + {\bar \psi} \eta ]} \nonumber \\
\label{zfqvaanc}
\eea
where
\bea
n^\mu =(1,0,0,0).
\eea
The correlation function of the type $<0|{\bar \psi}(x_2)\psi(x_1)|0>$ in QED in general Coulomb gauge is given by
\bea
&&<0|{\bar \psi}(x_2)\psi(x_1)|0>=\int [dQ] [d{\bar \psi}] [d \psi ]~{\bar \psi}(x_2)\psi(x_1) \nonumber \\
&&e^{i\int d^4x [-\frac{1}{4}{F}_{\mu \nu}^2[Q] -\frac{1}{2 \alpha} ([g^{\mu \nu}-\frac{n^\mu n^\nu}{n^2}]\partial_\mu Q_\nu)^2+{\bar \psi} [i\gamma^\mu \partial_\mu -m +e\gamma^\mu Q_\mu]\psi]}.
\label{zfqvcanc}
\eea

The generating functional in the background field method of QED in general Coulomb gauge is given by
\bea
Z[A,J,\eta,{\bar \eta}]=\int [dQ] [d{\bar \psi}] [d \psi ]~
e^{i\int d^4x [-\frac{1}{4}{F}_{\mu \nu}^2[Q] -\frac{1}{2 \alpha} ([g^{\mu \nu}-\frac{n^\mu n^\nu}{n^2}]\partial_\mu Q_\nu)^2+{\bar \psi}  [i\gamma^\mu \partial_\mu -m +e\gamma^\mu (A+Q)_\mu]  \psi+ J \cdot Q +{\bar \eta} \psi + {\bar \psi}\eta  ]}.\nonumber \\
\label{zfqiaxznc}
\eea
The correlation function of the type $<0|{\bar \psi}(x_2)\psi(x_1)|0>$ in the background field method of QED in general Coulomb gauge is given by
\bea
&&<0|{\bar \psi}(x_2)\psi(x_1)|0>_A=\int [dQ] [d{\bar \psi}] [d \psi ]~{\bar \psi}(x_2)\psi(x_1) \nonumber \\
&&e^{i\int d^4x [-\frac{1}{4}{F}_{\mu \nu}^2[Q] -\frac{1}{2 \alpha} ([g^{\mu \nu}-\frac{n^\mu n^\nu}{n^2}]\partial_\mu Q_\nu)^2+{\bar \psi}  [i\gamma^\mu \partial_\mu -m +e\gamma^\mu (A+Q)_\mu]  \psi]}.
\label{zfqvcbanc}
\eea
Hence by replacing $\frac{\eta^\mu \eta^\nu}{\eta^2} \rightarrow [g^{\mu \nu}-\frac{n^\mu n^\nu}{n^2}]$ everywhere in the derivations
in the previous sub-section we find
\bea
<{\bar \psi}(x_2) \psi(x_1)>=<{\bar \psi}(x_2)~e^{-ie\int_0^{\infty} d\lambda l\cdot { A}(x_2+l\lambda)}e^{ie\int_0^{\infty} d\lambda l\cdot { A}(x_1+l\lambda)} \psi(x_1)>_A
\label{tuccixanc}
\eea
which proves factorization of soft-collinear divergences in QED at all order in coupling constant in general Coulomb
gauge when light-like Wilson line is used.

\section{ Violation of Factorization Theorem of Soft-Collinear Divergences in QED With non-light-like Wilson line }

The main reason why factorization theorem of soft-collinear divergences in QED is violated when the non-light-like Wilson
line is used can be easily seen from eq. (\ref{jjj}) which is only valid for the light-like Wilson line but not valid for
the non-light-like Wilson line. When the Wilson line is non-light-like then in addition to $A_\mu(x)$ [like in the
right hand side of eq. (\ref{jjj})] one will also have a term containing non-zero $F_{\mu \nu}(x)$, the form of which
can be arbitrary, which spoils the factorization theorem.

For the non-light-like Wilson line we find
\bea
\partial_\mu [\int_0^x dx'^\nu A_\nu(x')]=\frac{\partial }{\partial x^\mu} [\int_0^1 d\tau \frac{dx'^\nu(\tau)}{d\tau} A_\nu(x'(\tau))],~~~~x'(0)=0,~~~~x'(1)=x
\eea
which gives
\bea
\partial_\mu [\int_0^x dx'^\nu A_\nu(x')]=\int_0^1 d\tau [ \frac{d[\frac{\partial x'^\nu(\tau)}{\partial x^\mu}]}{d\tau} A_\nu(x'(\tau))+\frac{\partial A_\nu(x'(\tau))}{\partial x'^\delta(\tau)}\frac{\partial x'^\delta(\tau)}{\partial x^\mu} \frac{dx'^\nu(\tau)}{d\tau} ].
\eea
Integrating by parts we find
\bea
&&\partial_\mu [\int_0^x dx'^\nu A_\nu(x')]=[\frac{\partial x'^\nu(\tau)}{\partial x^\mu} A_\nu(x'(\tau))]_{\tau=0}^{\tau=1}-\int_0^1 d\tau  \frac{\partial x'^\nu(\tau)}{\partial x^\mu} \frac{\partial A_\nu(x'(\tau))}{\partial x'^\delta(\tau)} \frac{dx'^\delta(\tau)}{d\tau}\nonumber \\
&&+\int_0^1 d\tau \frac{\partial A_\nu(x'(\tau))}{\partial x'^\delta(\tau)}\frac{\partial x'^\delta(\tau)}{\partial x^\mu} \frac{dx'^\nu(\tau)}{d\tau}
\eea
which gives
\bea
&&\partial_\mu [\int_0^x dx'^\nu A_\nu(x')]= A_\mu(x)+\int_0^1 d\tau F_{\nu \delta}(x'(\tau))
\frac{\partial x'^\delta(\tau)}{\partial x^\mu} \frac{dx'^\nu(\tau)}{d\tau}.
\label{jjh}
\eea

Note that for pure gauge we have
\bea
F_{\mu \nu}(x)=0
\label{o}
\eea
which when used in eq. in (\ref{jjh}) reproduces eq. (\ref{jjj}) which is valid for the light-like Wilson line.
Hence one finds that for the non-light-like Wilson line, it is the term containing non-zero $F_{\mu \nu}(x)$ in
the right hand side of eq. (\ref{jjh}), the form of which can be arbitrary, which spoils the factorization theorem
of soft-collinear divergences in QED.

\subsection{ Violation of Factorization Theorem in QED With Non-Light-Like Wilson Line in Covariant Gauge }

Note that eq. (\ref{zfqvcb}) is valid for any arbitrary background field $A^{\mu }(x)$. Hence we
start from eq. (\ref{zfqvcb}) by following exactly the same procedure that was followed for light-like
Wilson line case. By changing the integration variable from unprimed variable to primed variable we find from
eq. (\ref{zfqvcb})
\bea
<0|{\bar \psi}(x_2)\psi(x_1)|0>_A=\int [dQ] [d{\bar \psi}'] [d \psi' ]~{\bar \psi}'(x_2)\psi'(x_1)
e^{i\int d^4x [-\frac{1}{4}{F}_{\mu \nu}^2[Q] -\frac{1}{2 \alpha} (\partial_\mu Q^{\mu })^2+{\bar \psi}' [i\gamma^\mu \partial_\mu -m +e\gamma^\mu (A+Q)_\mu]  \psi']}.\nonumber \\
\label{zfqian}
\eea
This is because a change of integration variable from unprimed to primed variable does not change the value of the integration.
For the non-light-like Wilson line let us define the primed variable in analogous to the light-like Wilson line case
[similar to eq. (\ref{ppg})]
\bea
\psi'(x) = e^{-ie \int_0^{\infty} d\lambda v\cdot { A}(x+v\lambda) }\psi(x)
\label{qqgnn}
\eea
where $v^\mu$ is the non-light-like four velocity and $A^\mu(x)$ is not the U(1) pure gauge. Using contour integration
we find
\bea
&&\int_0^{\infty} d\lambda v\cdot  A(x+v\lambda) = \int_0^{\infty} d\lambda  e^{\lambda v \cdot \partial}~v \cdot  A(x)  =\int_0^{\infty} d\lambda  e^{\lambda v \cdot \partial}~\int \frac{d^4k}{(2\pi)^4} e^{ik \cdot x} v \cdot  A(k)  \nonumber \\
&& = \int \frac{d^4k}{(2\pi)^4} \int_0^{\infty} d\lambda  e^{i\lambda v \cdot k }~e^{ik \cdot x} v \cdot  A(k) =- \int \frac{d^4k}{(2\pi)^4}  \frac{1}{iv \cdot k }~e^{ik \cdot x} v \cdot  A(k) =- \frac{1}{v \cdot \partial }~ v \cdot  A(x) \nonumber \\
\label{ppgon}
\eea
which gives [since $A^\mu(x) \neq \partial^\mu \omega(x)$]
\bea
\int_0^{\infty} d\lambda v\cdot  A(x+v\lambda) =-\partial_\mu [\frac{1}{v \cdot \partial }~ v \cdot  A(x)] \neq -A_\mu(x).
\label{ppgona}
\eea
Note that when $A^\mu(x)$ is the U(1) pure gauge [$A^\mu(x) = \partial^\mu \omega(x)$] we find from eq. (\ref{ppgondc}) that
$\partial_\mu [\int_0^{\infty} d\lambda v\cdot  A(x+v\lambda)] = - A_\mu(x)$ which precisely cancels the $A^\mu$ present
in the lagrangian density in the background field method of QED [see the derivation of eq. (\ref{zfqibx}) from eq. (\ref{zfqia})]. This is the
reason why the soft-collinear divergences are factorized in QED when $A^\mu(x)$ is the U(1) pure gauge. However,
when $A^\mu(x)$ is not the U(1) pure gauge then such a cancelation of the $A^\mu(x)$ in the lagrangian density in the background
field method of QED does not happen [see eq. (\ref{ppgona})] which is why the soft-collinear divergences are not factorized in
QED when $A^\mu(x)$ is not the U(1) pure gauge.

From eq. (\ref{ppgona}) we find for the non-light-like Wilson line [$A^\mu(x) \neq \partial^\mu \omega(x)$] that
\bea
&&e^{-ie \int_0^{\infty} d\lambda v\cdot  A(x+v\lambda) } \neq e^{ie\omega(x)}
\label{ppgoo}
\eea
where $A^\mu(x)$ is not the U(1) pure gauge.

Hence we find that the equality relation holds for light-like Wilson line, as in eq. (\ref{phas}) where $A^\mu(x)$ is the U(1) pure gauge,
whereas the equality relation does not hold for non-light-like Wilson line, as in eq. (\ref{ppgoo})
where $A^\mu(x)$ is not the U(1) pure gauge. By using eqs. (\ref{qqgnn}) and (\ref{ppgon}) in (\ref{zfqian}) we find
\bea
&& <0|{\bar \psi}(x_2)\psi(x_1)|0>_A=\int [dQ] [d{\bar \psi}] [d \psi ]~{\bar \psi}(x_2)e^{ie \int_0^{\infty} d\lambda v\cdot { A}(x_2+v\lambda) }e^{-ie \int_0^{\infty} d\lambda v\cdot { A}(x_1+v\lambda) }\psi(x_1)\nonumber \\
&&\times e^{i\int d^4x [-\frac{1}{4}{F}_{\mu \nu}^2[Q] -\frac{1}{2 \alpha} (\partial_\mu Q^{\mu })^2+{\bar \psi} [i\gamma^\mu \partial_\mu -m +e\gamma^\mu (A+Q)_\mu  -e\gamma^\mu \partial_\mu [ \frac{1}{v \cdot \partial }~ v \cdot  A(x)]]  \psi ]}
\label{zfqibxnl}
\eea
which gives
\bea
&& <0|{\bar \psi}(x_2)e^{-ie \int_0^{\infty} d\lambda v\cdot { A}(x_2+v\lambda) }e^{ie \int_0^{\infty} d\lambda v\cdot { A}(x_1+v\lambda) }\psi(x_1)|0>_A=\int [dQ] [d{\bar \psi}] [d \psi ]~{\bar \psi}(x_2)\psi(x_1)\nonumber \\
&&\times e^{i\int d^4x [-\frac{1}{4}{F}_{\mu \nu}^2[Q] -\frac{1}{2 \alpha} (\partial_\mu Q^{\mu })^2+{\bar \psi} [i\gamma^\mu \partial_\mu -m +e\gamma^\mu (A+Q)_\mu  -e\gamma^\mu \partial_\mu [ \frac{1}{v \cdot \partial }~ v \cdot  A(x)]]  \psi ]}.
\label{zfqibxnl1}
\eea
Hence from eqs.  (\ref{ppgona}), (\ref{zfqibxnl1}) and (\ref{zfqvc}) we find
\bea
<0|{\bar \psi}(x_2)\psi(x_1)|0>\neq <0|{\bar \psi}(x_2)e^{-ie \int_0^{\infty} d\lambda v\cdot { A}(x_2+v\lambda) }e^{ie \int_0^{\infty} d\lambda v\cdot { A}(x_1+v\lambda) }\psi(x_1)|0>_A
\label{viol}
\eea
which implies the violation of the factorization theorem for soft-collinear divergences in QED at all order in
coupling constant when the non-light-like Wilson line is used in covariant gauge.
Note that the equality relation holds for light-like Wilson line, as in eq. (\ref{tuccix}), proving the factorization theorem
of soft-collinear divergences in QED when light-like Wilson line is used.

Hence by using path integral formulation of QED we have proved that the factorization theorem is valid
for light-like Wilson line [see eq. (\ref{tuccix})] but the factorization theorem is not valid for non-light-like Wilson line
[see eq. (\ref{viol})].

Since QCD is obtained from QED by extending U(1) gauge group to SU(3) gauge group
we expect that the factorization theorem in QCD should be valid
for light-like Wilson line but the factorization theorem in QCD should not be valid for non-light-like Wilson line.

\subsection{ Violation of Factorization Theorem in QED With Non-Light-Like Wilson Line in General Axial Gauge }

The correlation function of the type $<0|{\bar \psi}(x_2)\psi(x_1)|0>$ in the background field method of QED in general axial gauge is given by
eq. (\ref{zfqvcba}) which is valid for arbitrary background field $A^\mu(x)$. Hence from eq. (\ref{zfqvcba}) we find
\bea
<0|{\bar \psi}(x_2)\psi(x_1)|0>_A=\int [dQ] [d{\bar \psi}'] [d \psi' ]~{\bar \psi}'(x_2)\psi'(x_1)
e^{i\int d^4x [-\frac{1}{4}{F}_{\mu \nu}^2[Q] -\frac{1}{2 \alpha} (\eta_\mu Q^{\mu })^2+{\bar \psi}' [i\gamma^\mu \partial_\mu -m +e\gamma^\mu (A+Q)_\mu]  \psi']}\nonumber \\
\label{zfqiaav}
\eea
because the change of integration variable from unprimed to primed variable does not change the value of the integration.

By using eqs. (\ref{qqgnn}) and (\ref{ppgon}) in (\ref{zfqiaav}) we find
\bea
&& <0|{\bar \psi}(x_2)e^{-ie \int_0^{\infty} d\lambda v\cdot { A}(x_2+v\lambda) }e^{ie \int_0^{\infty} d\lambda v\cdot { A}(x_1+v\lambda) }\psi(x_1)|0>_A=\int [dQ] [d{\bar \psi}] [d \psi ]~{\bar \psi}(x_2)\psi(x_1)\nonumber \\
&&\times e^{i\int d^4x [-\frac{1}{4}{F}_{\mu \nu}^2[Q] -\frac{1}{2 \alpha} (\eta_\mu Q^{\mu })^2+{\bar \psi} [i\gamma^\mu \partial_\mu -m +e\gamma^\mu (A+Q)_\mu  -e\gamma^\mu \partial_\mu [ \frac{1}{v \cdot \partial }~ v \cdot  A(x)]]  \psi ]}.
\label{zfqibxnl1v}
\eea
Hence from eqs.  (\ref{ppgona}), (\ref{zfqibxnl1v}) and (\ref{zfqvca}) we find
\bea
<0|{\bar \psi}(x_2)\psi(x_1)|0>\neq <0|{\bar \psi}(x_2)e^{-ie \int_0^{\infty} d\lambda v\cdot { A}(x_2+v\lambda) }e^{ie \int_0^{\infty} d\lambda v\cdot { A}(x_1+v\lambda) }\psi(x_1)|0>_A
\label{violv}
\eea
which implies the violation of the factorization theorem for soft-collinear divergences in QED at all order in
coupling constant when the non-light-like Wilson line is used in general axial gauge.

\subsection{ Violation of Factorization Theorem in QED With Non-Light-Like Wilson Line in Light Cone Gauge }

Since eq. (\ref{violv}) is valid for general axial gauges [see eq. (\ref{axial})]
it is also valid for light cone gauge ($\eta^2=0$). Hence from eq. (\ref{violv}) we find that in the light cone gauge
\bea
<0|{\bar \psi}(x_2)\psi(x_1)|0>\neq <0|{\bar \psi}(x_2)e^{-ie \int_0^{\infty} d\lambda v\cdot { A}(x_2+v\lambda) }e^{ie \int_0^{\infty} d\lambda v\cdot { A}(x_1+v\lambda) }\psi(x_1)|0>_A
\label{viollc}
\eea
which implies the violation of the factorization theorem for soft-collinear divergences in QED at all order in
coupling constant when the non-light-like Wilson line is used in light cone gauge.

\subsection{ Violation of Factorization Theorem in QED With Non-Light-Like Wilson Line in General Non-Covariant Gauge }

The correlation function of the type $<0|{\bar \psi}(x_2)\psi(x_1)|0>$ in the background field method of QED in general
non-covariant gauge is given by eq. (\ref{zfqvcban}) which is valid for arbitrary background field $A^\mu(x)$.
Hence from eq. (\ref{zfqvcban}) we find
\bea
&& <0|{\bar \psi}(x_2)\psi(x_1)|0>_A=\int [dQ] [d{\bar \psi}'] [d \psi' ]~{\bar \psi}'(x_2)\psi'(x_1)\nonumber \\
&& e^{i\int d^4x [-\frac{1}{4}{F}_{\mu \nu}^2[Q] -\frac{1}{2 \alpha} (\frac{\eta^\mu \eta^\nu}{\eta^2}\partial_\mu Q_\nu)^2+{\bar \psi}' [i\gamma^\mu \partial_\mu -m +e\gamma^\mu (A+Q)_\mu]  \psi']}
\label{zfqiaanv}
\eea
because the change of integration variable from unprimed to primed variable does not change the value of the integration.

By using eqs. (\ref{qqgnn}) and (\ref{ppgon}) in (\ref{zfqiaanv}) we find
\bea
&& <0|{\bar \psi}(x_2)e^{-ie \int_0^{\infty} d\lambda v\cdot { A}(x_2+v\lambda) }e^{ie \int_0^{\infty} d\lambda v\cdot { A}(x_1+v\lambda) }\psi(x_1)|0>_A=\int [dQ] [d{\bar \psi}] [d \psi ]~{\bar \psi}(x_2)\psi(x_1)\nonumber \\
&&\times e^{i\int d^4x [-\frac{1}{4}{F}_{\mu \nu}^2[Q] -\frac{1}{2 \alpha} (\frac{\eta^\mu \eta^\nu}{\eta^2}\partial_\mu Q_\nu)^2+{\bar \psi} [i\gamma^\mu \partial_\mu -m +e\gamma^\mu (A+Q)_\mu  -e\gamma^\mu \partial_\mu [ \frac{1}{v \cdot \partial }~ v \cdot  A(x)]]  \psi ]}.
\label{zfqibxnl1vn}
\eea
Hence from eqs.  (\ref{ppgona}), (\ref{zfqibxnl1vn}) and (\ref{zfqvcan}) we find
\bea
<0|{\bar \psi}(x_2)\psi(x_1)|0>\neq <0|{\bar \psi}(x_2)e^{-ie \int_0^{\infty} d\lambda v\cdot { A}(x_2+v\lambda) }e^{ie \int_0^{\infty} d\lambda v\cdot { A}(x_1+v\lambda) }\psi(x_1)|0>_A
\label{violvn}
\eea
which implies the violation of the factorization theorem for soft-collinear divergences in QED at all order in
coupling constant when the non-light-like Wilson line is used in general non-covariant gauge.

\subsection{ Violation of Factorization Theorem in QED With Non-Light-Like Wilson Line in General Coulomb Gauge }

The correlation function of the type $<0|{\bar \psi}(x_2)\psi(x_1)|0>$ in the background field method of QED in general Coulomb
gauge is given by eq. (\ref{zfqvcbanc}) which is valid for arbitrary background field $A^\mu(x)$.

Hence by replacing $\frac{\eta^\mu \eta^\nu}{\eta^2} \rightarrow [g^{\mu \nu}-\frac{n^\mu n^\nu}{n^2}]$ everywhere in the derivations
in the previous sub-section we find that in general Coulomb gauge
\bea
<0|{\bar \psi}(x_2)\psi(x_1)|0>\neq <0|{\bar \psi}(x_2)e^{-ie \int_0^{\infty} d\lambda v\cdot { A}(x_2+v\lambda) }e^{ie \int_0^{\infty} d\lambda v\cdot { A}(x_1+v\lambda) }\psi(x_1)|0>_A
\label{violvnco}
\eea
which implies the violation of the factorization theorem for soft-collinear divergences in QED at all order in
coupling constant when the non-light-like Wilson line is used in general Coulomb gauge.

\section{ Soft-Collinear Divergences and Light-Like Wilson Line in QCD }

The gauge transformation of the quark field in QCD is given by
\bea
\psi'(x)=e^{igT^a\omega^a(x)}\psi(x).
\label{phg3}
\eea
Hence one finds that the issue of gauge invariance and factorization of soft-collinear divergences in QCD can be
simultaneously explained if $\omega^a(x)$ can be related to the soft-collinear gluon field $ {A}^{\mu a}(x)$. This is
easily done by using Eikonal Feynman rules in QCD for soft-collinear divergences when light-like Wilson line
is used which can be seen as follows.

The Eikonal propagator times the Eikonal vertex for a gluon with four momentum $k^\mu$ interacting with a
quark moving with four momentum $p^\mu$ in the limit $k<<<  p$ is given by
\cite{collins,tucci,berger,frederix,nayakqed,scet1,pathorder,nayaka2,nayaka3}
\bea
gT^a~\frac{p^\mu}{p \cdot k+i\epsilon }.
\label{eikonalinp}
\eea
In QCD the soft divergence arises only from the
emission of a gluon for which all components of the four-momentum $k^\mu$ are small
($k \rightarrow 0$) which is evident from eq. (\ref{eikonalinp}). From eq. (\ref{eikonalinp})
one also finds that when $0<k<<<  p$ and ${\vec p}$ is parallel to ${\vec k}$ one may
find collinear divergence in QCD.

However, since
\bea
p \cdot k =p_0 k_0 -{\vec p} \cdot {\vec k}=|{\vec p}||{\vec k}|(\sqrt{1+\frac{m^2}{{\vec p}^2}}  -{\rm cos}\theta)
\label{scdq}
\eea
one finds that the collinear divergence does not appear when quark interacts with a collinear gluon
because of the non-vanishing mass of the quark, {\it i. e.}, $m\neq 0$ even if the mass of the light quark is
very small. From eq. (\ref{scdq}) one finds that the collinear divergence appears only
when $m=0$ and $\theta =0$ where $\theta$ is the angle between ${\vec p}$ and ${\vec k}$.
Since gluons are massless and the massless gluons interact with each other one finds that the collinear
divergence appears in QCD. Since a massless particle is always light-like one
finds that the soft-collinear divergences can be described by light-like Wilson line in QCD.

For light-like quark we find from eq. (\ref{eikonalinp})
\bea
gT^a~\frac{p^\mu}{p \cdot k+i\epsilon }=gT^a~\frac{l^\mu}{l \cdot k+ i\epsilon }
\label{eikonalin}
\eea
where $l^\mu$ is the light-like four-velocity ($|{\vec l}|$=1) of the quark.
Note that when we say the "light-like quark" we mean the quark
that is traveling at its highest speed which is arbitrarily close to the speed of light
($|{\vec l}|\sim 1$) as it can not travel exactly at speed of light
because it has finite mass even if the mass of the light quark is very small.
On the other hand a massless gluon is light-like and it always remains light-like. Hence we find that
if $l^\mu$ is light-like four velocity then the soft-collinear divergence in QCD can be described by
the Eikonal Feynman rule as given by eq. (\ref{eikonalin}). Note that the Eikonal Feynman rule in eq.
(\ref{eikonalin}) is also valid if we replace the light-like quark by light-like gluon provided we replace
$T^{a}_{bc}=-if^{abc}$.

From eq. (\ref{eikonalin}) we find
\bea
&& gT^a\int \frac{d^4k}{(2\pi)^4} \frac{l\cdot { A}^a(k)}{l\cdot k +i\epsilon } =-gT^a i\int_0^{\infty} d\lambda \int \frac{d^4k}{(2\pi)^4} e^{i l \cdot k \lambda} l\cdot { A}^a(k) = igT^a\int_0^{\infty} d\lambda l\cdot {  A}^a(l\lambda)\nonumber \\
\label{ftgt}
\eea
where the gluon field $ { A}^{\mu a}(x)$ and its Fourier transform $ { A}^{\mu a}(k)$ are related by
\bea
{ A}^{\mu a}(x) =\int \frac{d^4k}{(2\pi)^4} { A}^{\mu a}(k) e^{ik \cdot x}.
\label{ft}
\eea
Note that a path ordering in QCD is required which can be seen as follows, see also \cite{bodwin}. The
Eikonal contribution for the soft-collinear divergence in QCD arising from a single soft-collinear gluon exchange in Feynman diagram
is given by eq. (\ref{ftgt}). Now consider the corresponding Feynman diagram for the soft-collinear divergences in QCD
due to exchange of two soft-collinear gluons of four-momenta $k^\mu_1$ and $k^\mu_2$.
The corresponding Eikonal contribution due to two soft-collinear gluons exchange is analogously given by
\bea
&& g^2\int \frac{d^4k_1}{(2\pi)^4} \frac{d^4k_2}{(2\pi)^4} \frac{T^a l\cdot { A}^a(k_2)T^b l\cdot { A}^b(k_1)}{(l\cdot (k_1+k_2) +i \epsilon)(l\cdot k_1 +i \epsilon)} \nonumber \\
&&=g^2i^2 \int_0^{\infty}  d\lambda_2 \int_{\lambda_2}^{\infty} d\lambda_1 T^a l\cdot { A}^a(l\lambda_2) T^b l\cdot { A}^b(l\lambda_1)
\nonumber \\
&&= \frac{g^2i^2}{2!} {\cal P}\int_0^{\infty}  d\lambda_2 \int_0^{\infty}  d\lambda_1 T^a l\cdot { A}^a(l\lambda_2) T^b l\cdot { A}^b(l\lambda_1)
\eea
where ${\cal P}$ is  the path ordering.
Extending this calculation up to infinite number of soft-collinear gluons we find that the Eikonal contribution for the soft-collinear
divergences due to soft-collinear gluons exchange with the light-like quark in QCD is given by the path ordered exponential
\bea
{\cal P}~{\rm exp}[ig \int_0^{\infty} d\lambda l\cdot { A}^a(l\lambda)T^a ]
\label{iiij}
\eea
where $l^\mu$ is the light-like four velocity of the quark. The Wilson line in QCD is given by
\bea
{\cal P}e^{ig \int_{x_i}^{x_f} dx^\mu A_\mu^a(x)T^a }
\label{tts}
\eea
which is the solution of the equation \cite{lam}
\bea
\partial_\mu S(x)=igT^aA_\mu^a(x)S(x)
\eea
with initial condition
\bea
S(x_i)=1.
\eea
When $A^{\mu a}(x)=A^{\mu a}(l\lambda)$ as in eq. (\ref{iiij}) we find from eq. (\ref{tts}) that the light-like Wilson line in
QCD for soft-collinear divergences is given by \cite{stermanpath}
\bea
{\cal P}e^{ig \int_{x_i}^{x_f} dx^\mu A_\mu^a(x) T^a}=\left[{\cal P}e^{-ig \int_0^{\infty} d\lambda l\cdot { A}^a(x_f+l\lambda) T^a}\right]{\cal P}e^{ig \int_0^{\infty} d\lambda l\cdot { A}^b(x_i+l\lambda) T^b}.
\label{oh}
\eea

A light-like quark traveling with light-like four-velocity $l^\mu$ produces SU(3) pure gauge potential $A^{\mu a}(x)$
at all the time-space position $x^\mu$ except at the position ${\vec x}$ perpendicular to the direction of motion
of the quark (${\vec l}\cdot {\vec x}=0$) at the time of closest approach \cite{stermanx,nayakj,nayake}.
When $A^{\mu a}(x) = A^{\mu a}(\lambda l)$ as in eq. (\ref{iiij})
we find ${\vec l}\cdot {\vec x}=\lambda {\vec l}\cdot {\vec l}=\lambda\neq 0$ which implies that the light-like Wilson line
finds the gluon field $A^{\mu a}(x)$ in eq. (\ref{iiij}) as the SU(3) pure gauge. The SU(3) pure gauge is given by
\bea
T^aA_\mu^a (x)= \frac{1}{ig}[\partial_\mu U(x)] ~U^{-1}(x),~~~~~~~~~~~~~U(x)=e^{igT^a\omega^a(x)}
\label{gtqcd}
\eea
which gives
\bea
U(x_f)={\cal P}e^{ig \int_{x_i}^{x_f} dx^\mu A_\mu^a(x) T^a}U(x_i)=e^{igT^a\omega^a(x_f)}.
\label{uxf}
\eea
Hence when $A^{\mu a}(x) = A^{\mu a}(\lambda l)$ as in eq. (\ref{iiij}) we find from eqs. (\ref{oh}) and
(\ref{uxf}) that the light-like Wilson line in QCD for soft-collinear divergences is given by
\bea
{\cal P}e^{ie \int_{x_i}^{x_f} dx^\mu A_\mu^a(x)T^a }=e^{igT^a\omega(x_f)}e^{-igT^b\omega^b(x_i)}=\left[{\cal P}e^{-ig \int_{0}^{\infty} d\lambda l\cdot { A}^a(x_f+l\lambda)T^a }\right]{\cal P}e^{ig \int_{0}^{\infty} d\lambda v\cdot { A}^b(x_i+l\lambda)T^b }\nonumber \\
\label{lkjn}
\eea
which depends only on end points $x_i^\mu$ and $x_f^\mu$ but is independent of the path.
The path independence can also be found from the non-abelian Stokes theorem which can be seen as follows.
The SU(3) pure gauge in eq. (\ref{gtqcd})
gives
\bea
F^a_{\mu \nu}[A]=\partial_\mu A^a_\nu(x) - \partial_\nu A^a_\mu(x)+gf^{abc} A^b_\mu(x) A^c_\nu(x)=0.
\label{cfmn}
\eea
Using eq. (\ref{cfmn}) in the non-abelian Stokes theorem \cite{stokes} we find
\bea
{\cal P}e^{ig \oint_C dx^\mu A_\mu^a(x)T^a} = {\cal P}{\rm exp}[ig \int_S dx^\mu dx^\nu \left[{\cal P}e^{ig \int_y^x dx'^\lambda A_\lambda^b(x')T^b}\right]F_{\mu \nu}^a(x)  T^a\left[{\cal P}e^{ig \int_x^y dx''^\delta A_\delta^c(x'')T^c}\right]]=1\nonumber \\
\label{555}
\eea
where $C$ is a closed path and $S$ is the surface enclosing $C$. Now considering two different paths
$L$ and $M$ with common end points $x_i^\mu$ and $x_f^\mu$ we find from eq. (\ref{555})
\bea
&&{\cal P}e^{ig \oint_C dx^\mu A_\mu^a(x)T^a} ={\cal P}{\rm exp}[ig \int_L dx^\mu A_\mu^a(x)T^a- ig \int_M dx^\mu A_\mu^a(x)T^a]\nonumber \\
&&=\left[{\cal P}e^{ig \int_L dx^\mu A_\mu^a(x)T^a}\right]\left[{\cal P}e^{- ig \int_M dx^\nu A_\nu^b(x)T^b}\right]=1
\label{pht}
\eea
which implies that the light-like Wilson line in QCD
\bea
{\cal P}e^{ig \int_{x_i}^{x_f} dx^\mu A_\mu^a(x)T^a }
\label{llwl}
\eea
depends only on the end points $x_i^\mu$ and $x_f^\mu$ but is independent of the path which can also be seen from eq. (\ref{lkjn}).
Hence from eq. (\ref{lkjn}) we find that the non-abelian phase or the gauge link in QCD is given by
\bea
\Phi(x)={\cal P}e^{-ig \int_0^{\infty} d\lambda l\cdot { A}^a(x+l\lambda)T^a }=e^{igT^a\omega^a(x)}.
\label{ttt}
\eea

Note that from eq. (\ref{cfmn}) we find the vanishing physical gauge invariant field strength square $F^{\mu \nu a}[A]F^a_{\mu \nu}[A]$
when $A^{\mu a}(x)$ is the SU(3) pure gauge as given by eq. (\ref{gtqcd}).
Hence in classical mechanics the SU(3) pure gauge potential does not have an effect on color charged
particle and one expects the effect of exchange of soft-collinear gluons to simply vanish.
However, in quantum mechanics the situation is a little more complicated, because the gauge potential
does have an effect on color charged particle even if it is SU(3) pure gauge potential and hence
one should not expect the effect of exchange of soft-collinear gluons to simply vanish \cite{stermanx}.
This can be verified by studying the non-perturbative correlation function
of the type $<0|{\bar \psi}(x) \psi(x') {\bar \psi}(x'') \psi(x''')...|0>$ in QCD in the
presence of SU(3) pure gauge background field.

Under non-abelian gauge transformation given by
\bea
T^aA'^a_\mu(x) = U(x)T^aA^a_\mu(x) U^{-1}(x)+\frac{1}{ig}[\partial_\mu U(x)] U^{-1}(x),~~~~~~~~~~~U(x)=e^{igT^a\omega^a(x)}
\label{aftgrmpi}
\eea
the Wilson line in QCD transforms as \cite{nayakj3}
\bea
{\cal P}e^{ie \int_{x_i}^{x_f} dx^\mu A'^a_\mu(x)T^a }=U(x_f)\left[{\cal P}e^{ie \int_{x_i}^{x_f} dx^\mu A^a_\mu(x)T^a }\right]U^{-1}(x_i).
\label{ty}
\eea
From eqs. (\ref{lkjn}) and (\ref{ty}) we find
\bea
{\cal P}e^{-ig \int_0^{\infty} d\lambda l\cdot { A}'^a(x+l\lambda)T^a }=U(x){\cal P}e^{-ig \int_0^{\infty} d\lambda l\cdot { A}^a(x+l\lambda)T^a },~~~~~~~~~~~~~~U(x)={\rm exp}[igT^a\omega^a(x)]\nonumber \\
\label{kkkj}
\eea
which gives from eq. (\ref{ttt})
\bea
\Phi'(x)=U(x)\Phi(x),~~~~~~~~~~~~~~~~~~~~~\Phi'^\dagger(x)=\Phi^\dagger(x)U^{-1}(x).
\label{hqg}
\eea

To summarize this, we find that the soft-collinear divergences in the perturbative Feynman diagrams due to soft-collinear gluons interaction
with the light-like Wilson line in QCD is given by the path ordered exponential in eq. (\ref{iiij}) which is nothing
but the non-abelian phase or the gauge link in QCD as given by eq. (\ref{ttt}) where the gluon field $A^{\mu a}(x)$ is the
SU(3) pure gauge, see eqs. (\ref{gtqcd}), (\ref{uxf}), (\ref{lkjn}). This implies that the effect of soft-collinear gluons
interaction between the partons and the light-like Wilson line in QCD can be studied by
putting the partons in the SU(3) pure gauge background field.
Hence we find that the soft-collinear behavior of the non-perturbative correlation function
of the type $<0|{\bar \psi}(x) \psi(x') {\bar \psi}(x'') \psi(x''')...|0>$ in QCD
due to the presence of light-like Wilson line in QCD can be studied by using the path integral method
of the QCD in the presence of SU(3) pure gauge background field.

It can be mentioned here that in soft collinear effective theory
(SCET) \cite{scet} it is also necessary to use the idea of background fields \cite{abbott}
to give well defined meaning to several distinct gluon fields \cite{scet1}.

Note that a massive color source traveling at speed much less than speed of light
can not produce SU(3) pure gauge field \cite{stermanx,nayakj,nayake}. Hence when one replaces light-like
Wilson line with massive Wilson line one expects the factorization of soft/infrared divergences to
break down. This is in conformation with the finding in \cite{nayaksterman,nayaksterman1} which used the diagrammatic
method of QCD to prove factorization theorem in NRQCD heavy quarkonium production in case of light-like Wilson line
\cite{nayaksterman} and violation of factorization theorem in NRQCD heavy quarkonium production in case of
non-light-like Wilson line \cite{nayaksterman1}. In case of massive Wilson line in QCD the color transfer occurs
and the factorization breaks down \cite{nayaksterman1}. Note that in case of massive Wilson line there is no
collinear divergences which is explained in eq. (\ref{scdq}).

\section{Proof of Factorization of Soft-Collinear Divergences in QCD With Light-Like Wilson Line }

By using path integral formulation of QED we have proved in section IV that the factorization theorem is valid
in QED when light-like Wilson line is used. Similarly, by using path integral formulation of QED we have
proved in section V that the factorization theorem is not valid in QED when non-light-like Wilson line is used.
Since QCD is obtained from QED by extending U(1) gauge group to SU(3) gauge group
we expect that the factorization theorem in QCD should be valid
for light-like Wilson line but the factorization theorem in QCD should not be valid for non-light-like Wilson line.

In this section we will prove that the Factorization theorem is valid in QCD for light-like Wilson line
and in the next section we will prove that the factorization theorem is not valid in QCD for
non-light-like Wilson line.

\subsection{ Proof of Factorization Theorem in QCD With Light-Like Wilson Line in Covariant Gauge }

The generating functional in the path integral method of QCD in the covariant gauge is given by \cite{muta,abbott}
\bea
Z[J,\eta,{\bar \eta}]=\int [dQ] [d{\bar \psi}] [d \psi ] ~{\rm det}(\frac{\delta (\partial_\mu Q^{\mu a})}{\delta \omega^b})
~e^{i\int d^4x [-\frac{1}{4}{F^a}_{\mu \nu}^2[Q] -\frac{1}{2 \alpha} (\partial_\mu Q^{\mu a})^2+{\bar \psi} [i\gamma^\mu \partial_\mu -m +gT^a\gamma^\mu Q^a_\mu] \psi + J \cdot Q +{\bar \eta} \psi +  {\bar \psi} \eta]} \nonumber \\
\label{zfq}
\eea
where $J^{\mu a}(x)$ is the external source for the quantum gluon field $Q^{\mu a}(x)$ and ${\bar \eta}_i(x)$ is the external source for the
Dirac field $\psi_i(x)$ of the quark and
\bea
F^a_{\mu \nu}[Q]=\partial_\mu Q^a_\nu(x)-\partial_\nu Q^a_\mu(x)+gf^{abc}Q^b_\mu(x)Q^c_\nu(x),~~~~~~~~~{F^a}_{\mu \nu}^2[Q]={F}^{\mu \nu a}[Q]{F}^a_{\mu \nu}[Q].
\eea
Note that the Faddeev-Popov (F-P) determinant ${\rm det}(\frac{\delta (\partial_\mu Q^{\mu a})}{\delta \omega^b})$ can be expressed in terms of path
integral over the ghost fields \cite{muta} but we will directly work with the Faddeev-Popov (F-P) determinant
${\rm det}(\frac{\delta (\partial_\mu Q^{\mu a})}{\delta \omega^b})$ in this paper.
The non-perturbative correlation function of the type $<0|{\bar \psi}(x_2) \psi(x_1)|0>$ in QCD in the covariant gauge is given by
\bea
&&<0|{\bar \psi}(x_2) \psi(x_1)|0>=\int [dQ] [d{\bar \psi}] [d \psi ] ~{\bar \psi}(x_2) \psi(x_1)\nonumber \\
&& \times {\rm det}(\frac{\delta (\partial_\mu Q^{\mu a})}{\delta \omega^b})~
e^{i\int d^4x [-\frac{1}{4}{F^a}_{\mu \nu}^2[Q] -\frac{1}{2 \alpha}(\partial_\mu Q^{\mu a})^2+{\bar \psi} [i\gamma^\mu \partial_\mu -m +gT^a\gamma^\mu Q^a_\mu] \psi  ]}.
\label{corq}
\eea

We have seen in the previous section that the soft-collinear behavior of the non-perturbative correlation function
due to the presence of light-like Wilson line in QCD can be studied by using the path integral method
of the QCD in the presence of SU(3) pure gauge background field. Hence we use the
path integral formulation of the background field method of QCD
in the presence of SU(3) pure gauge background field as given by eq. (\ref{gtqcd})
to prove factorization theorem in QCD.

Background field method of QCD was originally formulated by 't Hooft \cite{thooft} and later
extended by Klueberg-Stern and Zuber \cite{zuber,zuber1} and by Abbott \cite{abbott}.
This is an elegant formalism which can be useful to construct gauge invariant
non-perturbative green's functions in QCD. This formalism is also useful to study quark and gluon production from classical chromo field \cite{peter}
via Schwinger mechanism \cite{schw}, to compute $\beta$ function in QCD \cite{peskin}, to perform
calculations in lattice gauge theories \cite{lattice} and to study evolution of QCD
coupling constant in the presence of chromofield \cite{nayak}.

The generating functional in the path integral formulation of the background field method of QCD in the covariant gauge is
given by \cite{thooft,abbott,zuber}
\bea
&& Z[A,J,\eta,{\bar \eta}]=\int [dQ] [d{\bar \psi}] [d \psi ] ~{\rm det}(\frac{\delta G^a(Q)}{\delta \omega^b}) \nonumber \\
&& e^{i\int d^4x [-\frac{1}{4}{F^a}_{\mu \nu}^2[A+Q] -\frac{1}{2 \alpha}
(G^a(Q))^2+{\bar \psi} [i\gamma^\mu \partial_\mu -m +gT^a\gamma^\mu (A+Q)^a_\mu] \psi + J \cdot Q +{\bar \eta} \psi + {\bar \psi}\eta ]}
\label{azaqcd}
\eea
where the gauge fixing term is given by
\bea
G^a(Q) =\partial_\mu Q^{\mu a} + gf^{abc} A_\mu^b Q^{\mu c}=D_\mu[A]Q^{\mu a}
\label{ga}
\eea
which depends on the background field $A^{\mu a}(x)$ and
\bea
F_{\mu \nu}^a[A+Q]=\partial_\mu [A_\nu^a+Q_\nu^a]-\partial_\nu [A_\mu^a+Q_\mu^a]+gf^{abc} [A_\mu^b+Q_\mu^b][A_\nu^c+Q_\nu^c].
\eea
We have followed the notations of \cite{thooft,zuber,abbott} and accordingly we have
denoted the quantum gluon field by $Q^{\mu a}$ and the background field by $A^{\mu a}$.
Note that in the absence of the external sources the SU(3) pure gauge can be gauged away from the generating functional in the background
field method of QCD. However, in the presence of the external sources the SU(3) pure gauge can not be gauged away from the generating
functional in the background field method of QCD.

The non-perturbative correlation function
of the type $<0|{\bar \psi}(x_2) \psi(x_1)|0>_A$ in the background field method of QCD in the covariant gauge is given by
\bea
&&<0|{\bar \psi}(x_2) \psi(x_1)|0>_A=\int [dQ] [d{\bar \psi}] [d \psi ] ~{\bar \psi}(x_2) \psi(x_1)\nonumber \\
&& \times {\rm det}(\frac{\delta G^a(Q)}{\delta \omega^b}) e^{i\int d^4x [-\frac{1}{4}{F^a}_{\mu \nu}^2[A+Q] -\frac{1}{2 \alpha}
(G^a(Q))^2+{\bar \psi} [i\gamma^\mu \partial_\mu -m +gT^a\gamma^\mu (A+Q)^a_\mu] \psi   ]}.
\label{corqa}
\eea
The gauge fixing term $\frac{1}{2 \alpha} (G^a(Q))^2$ in eq. (\ref{azaqcd}) [where $G^a(Q)$ is given by eq. (\ref{ga})]
is invariant for gauge transformation of $A_\mu^a$:
\bea
\delta A_\mu^a = gf^{abc}A_\mu^b\omega^c + \partial_\mu \omega^a,  ~~~~~~~({\rm type~ I ~transformation})
\label{typeI}
\eea
provided one also performs a homogeneous transformation of $Q_\mu^a$ \cite{zuber,abbott}:
\bea
\delta Q_\mu^a =gf^{abc}Q_\mu^b\omega^c.
\label{omega}
\eea
The gauge transformation of background field $A_\mu^a$ as given by eq. (\ref{typeI})
along with the homogeneous transformation of $Q_\mu^a$ in eq. (\ref{omega}) gives
\bea
\delta (A_\mu^a+Q_\mu^a) = gf^{abc}(A_\mu^b+Q_\mu^b)\omega^c + \partial_\mu \omega^a
\label{omegavbxn}
\eea
which leaves $-\frac{1}{4}{F^a}_{\mu \nu}^2[A+Q]$ invariant in eq. (\ref{azaqcd}).

For fixed $A_\mu^a$, {\it i.e.}, for
\bea
&&\delta A_\mu^a =0,  ~~~~~~~({\rm type~ II ~transformation})
\label{typeII}
\eea
the gauge transformation of $Q_\mu^a$ \cite{zuber,abbott}:
\bea
&&\delta Q_\mu^a = gf^{abc}(A_\mu^b + Q_\mu^b)\omega^c + \partial_\mu \omega^a
\label{omegaII}
\eea
gives eq. (\ref{omegavbxn}) which leaves $-\frac{1}{4}{F^a}_{\mu \nu}^2[A+Q]$ invariant in eq. (\ref{azaqcd}).
Hence whether we use type I transformation [eqs. (\ref{typeI}) and (\ref{omega})] or type II transformation
[eqs. (\ref{typeII}) and (\ref{omegaII})] we will obtain the same equation (\ref{cfq5p2}).

It is useful to remember that, unlike QED \cite{tucci}, finding an exact relation between the generating
functional $Z[J, \eta, {\bar \eta}]$ in QCD in eq. (\ref{zfq}) and the generating functional $Z[A, J, \eta, {\bar \eta}]$
in the background field method of QCD in eq. (\ref{azaqcd}) in the presence of SU(3) pure gauge background field is not easy.
The main difficulty is due to the gauge fixing terms which are different in both the cases. While the Lorentz (covariant) gauge fixing
term  $-\frac{1}{2 \alpha}(\partial_\mu Q^{\mu a})^2$ in eq. (\ref{zfq}) in QCD is independent of the background field
$A^{\mu a}(x)$, the background field gauge fixing term $-\frac{1}{2 \alpha}(G^a(Q))^2$ in eq. (\ref{azaqcd}) in the background field method
of QCD depends on the background field $A^{\mu a}(x)$ where $G^a(Q)$ is given by eq. (\ref{ga}) \cite{thooft,zuber,abbott}.
Hence in order to study non-perturbative correlation function
in the background field method of QCD in the presence of SU(3) pure gauge background
field we proceed as follows.

By changing the integration variable $Q \rightarrow Q-A$ in the right hand side of eq. (\ref{corqa}) we find
\bea
&& <0|{\bar \psi}(x_2)\psi(x_1)|0>_A= \int [dQ] [d{\bar \psi}] [d \psi ]  ~{\bar \psi}(x_2)\psi(x_1) \nonumber \\
&& \times {\rm det}(\frac{\delta G_f^a(Q)}{\delta \omega^b})~~
e^{i\int d^4x [-\frac{1}{4}{F^a}_{\mu \nu}^2[Q] -\frac{1}{2 \alpha} (G_f^a(Q))^2+{\bar \psi} [i\gamma^\mu \partial_\mu -m +gT^a\gamma^\mu Q^a_\mu] \psi  ]}
\label{cfqcd1}
\eea
where from eq. (\ref{ga}) we find
\bea
G_f^a(Q) =\partial_\mu Q^{\mu a} + gf^{abc} A_\mu^b Q^{\mu c} - \partial_\mu A^{\mu a}=D_\mu[A] Q^{\mu a} - \partial_\mu A^{\mu a}
\label{gfa}
\eea
and from eq. (\ref{omega}) [by using eq. (\ref{typeI})] we find
\bea
\delta Q_\mu^a = -gf^{abc}\omega^b Q_\mu^c + \partial_\mu \omega^a.
\label{theta}
\eea
The eqs. (\ref{cfqcd1}), (\ref{gfa}) and (\ref{theta}) can also be derived by using type II transformation
which can be seen as follows. By changing $Q \rightarrow Q-A$ in eq. (\ref{corqa}) we find eq. (\ref{cfqcd1})
where the gauge fixing term from eq. (\ref{ga}) becomes eq. (\ref{gfa})
and eq. (\ref{omegaII}) [by using eq. (\ref{typeII})] becomes eq. (\ref{theta}).
Hence we obtain eqs. (\ref{cfqcd1}), (\ref{gfa}) and (\ref{theta}) whether we use
the type I transformation or type II transformation. Hence we find that we will obtain the same
eq. (\ref{cfq5p2}) whether we use the type I transformation or type II transformation.

Changing the integration variable from unprimed variable to primed variable we find from eq. (\ref{cfqcd1})
\bea
&& <0|{\bar \psi}(x_2)\psi(x_1)|0>_A= \int [dQ'] [d{\bar \psi}'] [d \psi' ]~{\bar \psi}'(x_2)\psi'(x_1) \nonumber \\
&& \times {\rm det}(\frac{\delta G_f^a(Q')}{\delta \omega^b})~~
e^{i\int d^4x [-\frac{1}{4}{F^a}_{\mu \nu}^2[Q'] -\frac{1}{2 \alpha} (G_f^a(Q'))^2+{\bar \psi}' [i\gamma^\mu \partial_\mu -m +gT^a\gamma^\mu Q'^a_\mu] \psi'  ]}.
\label{cfqcd1vb}
\eea
This is because a change of integration variable from unprimed variable to primed variable does not change the value of the
integration.

The equation
\bea
Q'^a_\mu(x)= Q^a_\mu(x) +gf^{abc}\omega^c(x) Q_\mu^b(x) + \partial_\mu \omega^a(x)
\label{thetaav}
\eea
in eq. (\ref{theta}) is valid for infinitesimal transformation ($\omega << 1$) which is obtained from the
finite equation
\bea
&&T^aQ'^a_\mu(x) = U(x)T^aQ^a_\mu(x) U^{-1}(x)+\frac{1}{ig}[\partial_\mu U(x)] U^{-1}(x)
\label{ftgrm}
\eea
where
\bea
U(x)={\cal P}e^{-ig \int_0^{\infty} d\lambda l\cdot { A}^a(x+l\lambda)T^a }=e^{igT^a\omega^a(x)}.
\label{ftgrmu}
\eea
Simplifying infinite numbers of non-commuting terms we find
\bea
\left[~e^{-igT^b\omega^b(x)} ~T^a~ e^{igT^c \omega^c(x)}~\right]_{ij}=[e^{-gM(x)}]_{ab}T^b_{ij}
\label{non}
\eea
where
\bea
M_{ab}(x)=f^{abc}\omega^c(x).
\label{mab}
\eea
Hence by simplifying infinite numbers of non-commuting terms in eq. (\ref{ftgrm}) [by using eq. (\ref{non}) and \cite{nayakj}] we find that
\bea
{Q'}_\mu^a(x) = [e^{gM(x)}]_{ab}Q_\mu^b(x) ~+ ~[\frac{e^{gM(x)}-1}{gM(x)}]_{ab}~[\partial_\mu \omega^b(x)],~~~~~~~~~~~M_{ab}(x)=f^{abc}\omega^c(x).
\label{teq}
\eea

Under the finite transformation, using eq. (\ref{teq}), we find
\bea
&& [dQ'] =[dQ] ~{\rm det} [\frac{\partial {Q'}^a}{\partial Q^b}] = [dQ] ~{\rm det} [[e^{gM(x)}]]=[dQ] {\rm exp}[{\rm Tr}({\rm ln}[e^{gM(x)}])]=[dQ]
\label{dqa}
\eea
where we have used (for any matrix $H$)
\bea
{\rm det}H={\rm exp}[{\rm Tr}({\rm ln}H)].
\eea

Let us define
\bea
\psi'(x) = {\cal P}e^{-ig \int_0^{\infty} d\lambda l\cdot { A}^a(x+l\lambda)T^a }\psi(x)=e^{igT^a\omega^a(x)}\psi(x).
\label{ppgn}
\eea
When $A^{\mu a}(x)$ is SU(3) pure gauge we find by using eqs. (\ref{teq}) and (\ref{ppgn}) that
\bea
&&[d{\bar \psi}'] [d \psi' ]=[d{\bar \psi}] [d \psi ],~~~~~~{\bar \psi}' [i\gamma^\mu \partial_\mu -m +gT^a\gamma^\mu Q'^a_\mu] \psi'={\bar \psi} [i\gamma^\mu \partial_\mu -m +gT^a\gamma^\mu Q^a_\mu]\psi,\nonumber \\
&&{F^a}_{\mu \nu}^2[Q']={F^a}_{\mu \nu}^2[Q].
\label{psa}
\eea
Simplifying all the infinite number of non-commuting terms in eq. (\ref{gtqcd}) we find that the
SU(3) pure gauge $A^{\mu a}(x)$ is given by \cite{nayakj}
\bea
A^{\mu a}(x)=\partial^\mu \omega^b(x)\left[\frac{e^{gM(x)}-1}{gM(x)}\right]_{ab}
\label{pg4}
\eea
where $M_{ab}(x)$ is given by eq. (\ref{mab}). From eqs. (\ref{pg4}) and (\ref{teq}) we find
\bea
{Q'}_\mu^a(x) -A_\mu^a(x)= [e^{gM(x)}]_{ab}Q_\mu^b(x),~~~~~~~~~~~M_{ab}(x)=f^{abc}\omega^c(x).
\label{tej}
\eea

Using eqs. (\ref{dqa}), (\ref{psa}) and (\ref{tej}) in eq. (\ref{cfqcd1vb}) we find
\bea
&& <0|{\bar \psi}(x_2)\psi(x_1)|0>_A=~\int [dQ] [d{\bar \psi}] [d \psi ] {\bar \psi}(x_2)[{\cal P}e^{-igT^a \int_{x_1}^{x_2} dx^\mu A_\mu^a(x)}]\psi(x_1)~\nonumber \\
 && \times {\rm det}(\frac{\delta G_f^a(Q')}{\delta \omega^b})
~e^{i\int d^4x [-\frac{1}{4}{F^a}_{\mu \nu}^2[Q] -\frac{1}{2 \alpha} (G_f^a(Q'))^2+{\bar \psi} [i\gamma^\mu \partial_\mu -m +gT^a\gamma^\mu Q^a_\mu] \psi ]}.
\label{cfqcd1c}
\eea

From eq. (\ref{gfa}) we find
\bea
G_f^a(Q') =\partial_\mu Q^{' \mu a} + gf^{abc} A_\mu^b Q^{' \mu c} - \partial_\mu A^{\mu a}.
\label{gfap}
\eea
By using eqs. (\ref{teq}) and (\ref{pg4}) in eq. (\ref{gfap}) we find
\bea
&&G_f^a(Q') =\partial^\mu [[e^{gM(x)}]_{ab}Q_\mu^b(x) ~+ ~[\frac{e^{gM(x)}-1}{gM(x)}]_{ab}~[\partial_\mu \omega^b(x)]]\nonumber \\
&&+ gf^{abc} [\partial^\mu \omega^e(x)\left[\frac{e^{gM(x)}-1}{gM(x)}\right]_{be}] [[e^{gM(x)}]_{cd}Q_\mu^d(x) ~+ ~[\frac{e^{gM(x)}-1}{gM(x)}]_{cd}~[\partial_\mu \omega^d(x)]]\nonumber \\
&&- \partial_\mu [\partial^\mu \omega^b(x)\left[\frac{e^{gM(x)}-1}{gM(x)}\right]_{ab}]
\label{gfapa}
\eea
which gives
\bea
&&G_f^a(Q') =\partial^\mu [[e^{gM(x)}]_{ab}Q_\mu^b(x)]\nonumber \\
&&+ gf^{abc} [\partial^\mu \omega^e(x)\left[\frac{e^{gM(x)}-1}{gM(x)}\right]_{be}] [[e^{gM(x)}]_{cd}Q_\mu^d(x) ~+ ~[\frac{e^{gM(x)}-1}{gM(x)}]_{cd}~[\partial_\mu \omega^d(x)]].
\label{gfapb}
\eea
From eq. (\ref{gfapb}) we find
\bea
&&G_f^a(Q') =\partial^\mu [[e^{gM(x)}]_{ab}Q_\mu^b(x)]+ gf^{abc} [[\partial^\mu \omega^e(x)]\left[\frac{e^{gM(x)}-1}{gM(x)}\right]_{be}] [[e^{gM(x)}]_{cd}Q_\mu^d(x)]
\label{gfapc}
\eea
which gives
\bea
&&G_f^a(Q') = [e^{gM(x)}]_{ab}\partial^\mu Q_\mu^b(x)\nonumber \\
&&+Q_\mu^b(x)\partial^\mu [[e^{gM(x)}]_{ab}]+  [[\partial^\mu \omega^e(x)]\left[\frac{e^{gM(x)}-1}{gM(x)}\right]_{be}]gf^{abc} [[e^{gM(x)}]_{cd}Q_\mu^d(x)].
\label{gfapd}
\eea
From \cite{nayakj} we find
\bea
\partial^\mu [e^{igT^a\omega^a(x)}]_{ij}=ig[\partial^\mu \omega^b(x)]\left[\frac{e^{gM(x)}-1}{gM(x)}\right]_{ab}T^a_{ik}[e^{igT^c\omega^c(x)}]_{kj},~~~~~~~~~M_{ab}(x)=f^{abc}\omega^c(x)\nonumber \\
\label{pg4j}
\eea
which in the adjoint representation of SU(3) gives (by using $T^a_{bc}=-if^{abc}$)
\bea
[\partial^\mu e^{gM(x)}]_{ad}=[\partial^\mu \omega^e(x)]\left[\frac{e^{gM(x)}-1}{gM(x)}\right]_{be}gf^{bac}[e^{M(x)}]_{cd},~~~~~~~~~M_{ab}(x)=f^{abc}\omega^c(x).
\label{pg4k}
\eea
Using eq. (\ref{pg4k}) in (\ref{gfapd}) we find
\bea
&&G_f^a(Q') = [e^{gM(x)}]_{ab}\partial^\mu Q_\mu^b(x)
\label{gfape}
\eea
which gives
\bea
(G_f^a(Q'))^2 = (\partial_\mu Q^{\mu a}(x))^2.
\label{gfapf}
\eea
Since for $n \times n$ matrices $A$ and $B$ we have
\bea
{\rm det}(AB)=({\rm det}A)({\rm det} B)
\label{detr}
\eea
we find by using eq. (\ref{gfape}) that
\bea
&&{\rm det} [\frac{\delta G_f^a(Q')}{\delta \omega^b}] ={\rm det}
[\frac{ \delta [[e^{gM(x)}]_{ac}\partial^\mu Q_\mu^c(x)]}{\delta \omega^b}]={\rm det}[
[e^{gM(x)}]_{ac}\frac{ \delta (\partial^\mu Q_\mu^c(x))}{\delta \omega^b}]\nonumber \\
&&=\left[{\rm det}[
[e^{gM(x)}]_{ac}]\right]~\left[{\rm det}[\frac{ \delta (\partial^\mu Q_\mu^c(x))}{\delta \omega^b}]\right]={\rm exp}[{\rm Tr}({\rm ln}[e^{gM(x)}])]~{\rm det}[\frac{ \delta (\partial_\mu Q^{\mu a}(x))}{\delta \omega^b}]\nonumber \\
&&={\rm det}[\frac{ \delta (\partial_\mu Q^{\mu a}(x))}{\delta \omega^b}].
\label{gqp4a}
\eea
Using eqs. (\ref{gfapf}) and (\ref{gqp4a}) in eq. (\ref{cfqcd1c}) we find
\bea
&&<0|{\bar \psi}(x_2)\psi(x_1)|0>_A=\int [dQ] [d{\bar \psi}] [d \psi ] ~{\bar \psi}(x_2)[{\cal P}e^{-igT^a \int_{x_1}^{x_2} dx^\mu A_\mu^a(x)}]\psi(x_1)\nonumber \\
 && \times
{\rm det}(\frac{\delta (\partial_\mu Q^{\mu a})}{\delta \omega^b})~
e^{i\int d^4x [-\frac{1}{4}{F^a}_{\mu \nu}^2[Q] -\frac{1}{2 \alpha}(\partial_\mu Q^{\mu a})^2+{\bar \psi} [i\gamma^\mu \partial_\mu -m +gT^a\gamma^\mu Q^a_\mu] \psi  ]}.
\label{cfq5p1}
\eea
Using the similar technique as above we find
\bea
&&
<0|{\bar \psi}(x_2)[{\cal P}e^{igT^a \int_{x_1}^{x_2} dx^\mu A_\mu^a(x)}]\psi(x_1)|0>_A
 =\int [dQ] [d{\bar \psi}] [d \psi ] ~{\bar \psi}(x_2)\psi(x_1)\nonumber \\
 && \times {\rm det}(\frac{\delta (\partial_\mu Q^{\mu a})}{\delta \omega^b})~
e^{i\int d^4x [-\frac{1}{4}{F^a}_{\mu \nu}^2[Q] -\frac{1}{2 \alpha}(\partial_\mu Q^{\mu a})^2+{\bar \psi} [i\gamma^\mu \partial_\mu -m +gT^a\gamma^\mu Q^a_\mu] \psi  ]}
\label{cfq5p2}
\eea
in the presence of SU(3) pure gauge background field $A^{\mu a}(x)$ as given by eq. (\ref{gtqcd}) where $M_{ab}(x)$
is given by eq. (\ref{mab}).

Note that eq. (\ref{cfq5p2}) is valid whether we use type I transformation [eqs. (\ref{typeI}) and (\ref{omega})] or type II transformation
[eqs. (\ref{typeII}) and (\ref{omegaII})]. However, since eq. (\ref{aftgrmpi})
is used to study the gauge transformation of the Wilson line in
QCD as given by eq. (\ref{ty}), we will use type I transformation [see eqs. (\ref{typeI}) and (\ref{omega})] in the rest of the paper
which for the finite transformation give eq. (\ref{aftgrmpi}) and  \cite{abbott,zuber}
\bea
T^aQ'^a_\mu(x) = U(x)T^aQ^a_\mu(x) U^{-1}(x),~~~~~~~~~~~U(x)=e^{igT^a\omega^a(x)}.
\label{jprtpi}
\eea
 From eq.
(\ref{jprtpi}) we find that under gauge transformation given by eq. (\ref{aftgrmpi}) the (quantum) gluon field $Q^{\mu a}(x)$ transforms as
\bea
Q'^a_\mu(x)=[e^{gM(x)}]_{ab}Q_\mu^b(x)
\label{jpr}
\eea
where $M_{ab}(x)$ is given by eq. (\ref{mab}). From eqs. (\ref{corq}) and (\ref{cfq5p2}) we find
\bea
<0|{\bar \psi}(x_2)\psi(x_1)|0> =<0|{\bar \psi}(x_2)[{\cal P}e^{igT^a \int_{x_1}^{x_2} dx^\mu A_\mu^a(x)}]\psi(x_1)|0>_A
\label{finaly}
\eea
which proves factorization of soft-collinear divergences at all order in coupling constant in QCD when light-like Wilson line is
used in the covariant gauge where
\bea
&&{\cal P}e^{igT^a \int_{x_1}^{x_2} dx^\mu A_\mu^a(x)}=[{\cal P}{\rm exp}[-ig\int_0^{\infty} d\lambda l\cdot { A}^c(x_2+l\lambda)T^{c}]]\nonumber \\
&&\times {\cal P}{\rm exp}[ig\int_0^{\infty} d\lambda l\cdot { A}^b(x_1+l\lambda)T^{b}]
\label{wilabf}
\eea
is the light-like Wilson line in terms of the non-abelian gauge link or non-abelian phase and $l^\mu$ is the light-like four velocity.

\subsection{ Proof of Factorization Theorem in QCD With Light-Like Wilson Line in General Axial Gauge }

In QCD the generating functional with general axial gauge fixing is given by \cite{leib,meis}
\bea
&& Z[J,\eta,{\bar \eta}]=\int [dQ] [d{\bar \psi}] [d \psi ] \nonumber \\
&& \times e^{i\int d^4x [-\frac{1}{4}{F^a}_{\mu \nu}^2[Q] -\frac{1}{2 \alpha} (\eta_\mu Q^{\mu a})^2+{\bar \psi} [i\gamma^\mu \partial_\mu -m +gT^a\gamma^\mu Q^a_\mu]  \psi + J \cdot Q +{\bar \psi} \eta +{\bar \eta} \psi ]}
\label{zfqag}
\eea
where $\eta^\mu$ is an arbitrary but constant four vector
\bea
&&\eta^2 <0,~~~~~~~~{\rm axial~gauge}\nonumber \\
&& \eta^2 =0,~~~~~~~~{\rm light}-{\rm cone~gauge}\nonumber \\
&& \eta^2 >0,~~~~~~~~{\rm temporal~gauge}.
\label{axialg}
\eea
Note that unlike covariant gauge in eq. (\ref{zfq}) there is no Faddeev-Popov (F-P) determinant in eq. (\ref{zfqag}) because the
ghost particles decouple in general axial gauges \cite{leib,meis}.
The non-perturbative correlation function of the type $<0|{\bar \psi}(x_2) \psi(x_1)|0>$ in QCD in general axial gauge
is given by
\bea
&&<0|{\bar \psi}(x_2) \psi(x_1)|0>=\int [dQ] [d{\bar \psi}] [d \psi ] ~{\bar \psi}(x_2) \psi(x_1)\nonumber \\
&& \times
e^{i\int d^4x [-\frac{1}{4}{F^a}_{\mu \nu}^2[Q] -\frac{1}{2 \alpha} (\eta^\mu Q_\mu^a)^2+{\bar \psi} [i\gamma^\mu \partial_\mu -m +gT^a\gamma^\mu Q^a_\mu] \psi  ]}.
\label{cfq5ag}
\eea

The generating functional in the background field method of QCD with general axial gauge fixing is given by \cite{meis}
\bea
&& Z[A,J,\eta,{\bar \eta}]=\int [dQ] [d{\bar \psi}] [d \psi ] \nonumber \\
&& \times e^{i\int d^4x [-\frac{1}{4}{F^a}_{\mu \nu}^2[A+Q] -\frac{1}{2 \alpha}
(\eta^\mu Q_\mu^a)^2+{\bar \psi} [i\gamma^\mu \partial_\mu -m +gT^a\gamma^\mu (A+Q)^a_\mu] \psi + J \cdot Q +{\bar \eta} \psi +{\bar \psi} \eta  ]}.
\label{azaqcdag}
\eea

The non-perturbative correlation function of the type $<0|{\bar \psi}(x_2) \psi(x_1)|0>_A$
in the background field method of QCD in general axial gauges is given by
\bea
&&<0|{\bar \psi}(x_2) \psi(x_1)|0>_A=\int [dQ] [d{\bar \psi}] [d \psi ] ~{\bar \psi}(x_2) \psi(x_1)\nonumber \\
&& \times e^{i\int d^4x [-\frac{1}{4}{F^a}_{\mu \nu}^2[A+Q] -\frac{1}{2 \alpha}
(\eta^\mu Q_\mu^a)^2+{\bar \psi} [i\gamma^\mu \partial_\mu -m +gT^a\gamma^\mu (A+Q)^a_\mu] \psi   ]}.
\label{cfqcdag}
\eea

By changing the integration variable $Q \rightarrow Q-A$ in the right hand side of eq. (\ref{cfqcdag}) we find
\bea
&& <0|{\bar \psi}(x_2) \psi(x_1)|0>_A= \int [dQ] [d{\bar \psi}] [d \psi ]  ~{\bar \psi}(x_2)\psi(x_1)\nonumber \\
&& \times
e^{i\int d^4x [-\frac{1}{4}{F^a}_{\mu \nu}^2[Q] -\frac{1}{2 \alpha} (\eta^\mu (Q-A)_\mu^a)^2+{\bar \psi} [i\gamma^\mu \partial_\mu -m +gT^a\gamma^\mu Q^a_\mu] \psi  ]}.
\label{cfqcd1ag}
\eea

Changing the integration variable from unprimed variable to primed variable we find from eq. (\ref{cfqcd1ag})
\bea
&& <0|{\bar \psi}(x_2) \psi(x_1)|0>_A= \int [dQ'] [d{\bar \psi}'] [d \psi' ]~{\bar \psi}'(x_2)\psi'(x_1) \nonumber \\
&& \times
e^{i\int d^4x [-\frac{1}{4}{F^a}_{\mu \nu}^2[Q'] -\frac{1}{2 \alpha} (\eta^\mu (Q'-A)_\mu^a)^2+{\bar \psi}' [i\gamma^\mu \partial_\mu -m +gT^a\gamma^\mu Q'^a_\mu] \psi'  ]}.
\label{cfqcd1vbag}
\eea
This is because a change of integration variable from unprimed variable to primed variable does not change the value of the
integration.

Using eqs. (\ref{dqa}), (\ref{psa}) and (\ref{tej}) in eq. (\ref{cfqcd1vbag}) we find
\bea
&& <0|{\bar \psi}(x_2) \psi(x_1)|0>_A=~\int [dQ] [d{\bar \psi}] [d \psi ] ~{\bar \psi}(x_2)[{\cal P}e^{-igT^a \int_{x_1}^{x_2} dx^\mu A_\mu^a(x)}]\psi(x_1)~\nonumber \\
 && \times e^{i\int d^4x [-\frac{1}{4}{F^a}_{\mu \nu}^2[Q] -\frac{1}{2 \alpha} (\eta^\mu ([e^{gM(x)}]_{ab}Q_\mu^b(x)))^2+{\bar \psi} [i\gamma^\mu \partial_\mu -m +gT^a\gamma^\mu Q^a_\mu] \psi ]}
\label{cfqcd1caig}
\eea
which gives
\bea
&& <0|{\bar \psi}(x_2) \psi(x_1)|0>_A=~\int [dQ] [d{\bar \psi}] [d \psi ] ~~{\bar \psi}(x_2)[{\cal P}e^{-igT^a \int_{x_1}^{x_2} dx^\mu A_\mu^a(x)}]\psi(x_2)~\nonumber \\
 && \times e^{i\int d^4x [-\frac{1}{4}{F^a}_{\mu \nu}^2[Q] -\frac{1}{2 \alpha} (\eta^\mu Q_\mu^a)^2+{\bar \psi} [i\gamma^\mu \partial_\mu -m +gT^a\gamma^\mu Q^a_\mu] \psi ]}.
\label{cfq5p1ag}
\eea

Using the similar technique as above we find
\bea
&&
<0|~{\bar \psi}(x_2)[{\cal P}e^{igT^a \int_{x_1}^{x_2} dx^\mu A_\mu^a(x)}]\psi(x_1)~|0>_A
 =\int [dQ] [d{\bar \psi}] [d \psi ] ~~{\bar \psi}(x_2)\psi(x_1)~\nonumber \\
 && \times
e^{i\int d^4x [-\frac{1}{4}{F^a}_{\mu \nu}^2[Q] -\frac{1}{2 \alpha}(\eta^\mu Q_\mu^a)^2+{\bar \psi} [i\gamma^\mu \partial_\mu -m +gT^a\gamma^\mu Q^a_\mu] \psi  ]}
\label{cfq5p2ag}
\eea
in general axial gauge
in the presence of SU(3) pure gauge background field $A^{\mu a}(x)$ as given by eq. (\ref{gtqcd}).

From eqs. (\ref{lkjn}), (\ref{cfq5ag}) and (\ref{cfq5p2ag}) we find
\bea
&&<0|{\bar \psi}(x_2) \psi(x_1)|0> =<0|{\bar \psi}(x_2) [{\cal P}{\rm exp}[-ig\int_0^{\infty} d\lambda l\cdot { A}^c(x_2+l\lambda)T^{c}]]\nonumber \\
&&\times {\cal P}{\rm exp}[ig\int_0^{\infty} d\lambda l\cdot { A}^b(x_1+l\lambda)T^{b}]\psi_(x_1)|0>_A
\label{finalzag}
\eea
which proves factorization of soft-collinear divergences at all order in coupling constant in QCD when light-like Wilson line is
used in general axial gauge.

\subsection{ Proof of Factorization Theorem in QCD With Light-Like Wilson Line in Light Cone Gauge }

The light-cone gauge corresponds to \cite{leib,meis,collinsp}
\bea
\eta \cdot Q^a=0,~~~~~~~~~~~~~~~~\eta^2=0
\label{jnkg}
\eea
which is already covered by eqs. (\ref{zfqag}) and (\ref{axialg}) where the corresponding gauge fixing term
is given by -$\frac{1}{2 \alpha} (\eta_\mu Q^{\mu a})^2$. In the light-cone coordinate system
the light-cone gauge \cite{collinsp}
\bea
Q^{+a}=0
\label{lightcg}
\eea
corresponds to
\bea
\eta^\mu = (\eta^+,\eta^-,\eta_\perp)= (0,1,0)
\label{ligtcg}
\eea
which covers $\eta \cdot Q^a=0$ and $\eta^2=0$ situation in eq. (\ref{jnkg}).

Since eq. (\ref{finalzag}) is valid for general axial gauge [see eq. (\ref{axialg})] it is also valid in light cone
gauge ($\eta^2=0$). Hence in light cone gauge we find from eq. (\ref{finalzag}) that
\bea
&&<0|{\bar \psi}(x_1) \psi(x_2)|0>=<0|{\bar \psi}(x_1) [{\cal P}{\rm exp}[-ig\int_0^{\infty} d\lambda l\cdot { A}^c(x_1+l\lambda)T^{c}]]\nonumber \\
&&\times {\cal P}{\rm exp}[ig\int_0^{\infty} d\lambda l\cdot { A}^b(x_2+l\lambda)T^{b}]\psi_(x_2)|0>_A
\label{finalzlg}
\eea
which proves factorization of soft-collinear divergences at all order in coupling constant in QCD when light-like Wilson line is
used in light-cone gauge.

\subsection{ Proof of Factorization Theorem in QCD With Light-Like Wilson Line in General Non-Covariant Gauge }

In QCD the generating functional with general non-covariant gauge fixing is given by \cite{noncov,noncov1}
\bea
&& Z[J,\eta,{\bar \eta}]=\int [dQ] [d{\bar \psi}] [d \psi ] ~{\rm det}(\frac{\delta (\frac{\eta^\mu \eta^\nu}{\eta^2}\partial_\mu Q_\nu^a)}{\delta \omega^b}) \nonumber \\
&& \times e^{i\int d^4x [-\frac{1}{4}{F^a}_{\mu \nu}^2[Q] -\frac{1}{2 \alpha} (\frac{\eta^\mu \eta^\nu}{\eta^2}\partial_\mu Q_\nu^a)^2+{\bar \psi} [i\gamma^\mu \partial_\mu -m +gT^a\gamma^\mu Q^a_\mu]  \psi + J \cdot Q +{\bar \psi} \eta +{\bar \eta} \psi ]}
\label{zfqng}
\eea
where $\eta^\mu$ is an arbitrary but constant four vector.
The non-perturbative correlation function of the type $<0|{\bar \psi}(x_2) \psi(x_1)|0>$ in QCD in general non-covariant gauge
is given by
\bea
&&<0|{\bar \psi}(x_2) \psi(x_1)|0>=\int [dQ] [d{\bar \psi}] [d \psi ] ~{\bar \psi}(x_2) \psi(x_1)\nonumber \\
&& \times ~{\rm det}(\frac{\delta (\frac{\eta^\mu \eta^\nu}{\eta^2}\partial_\mu Q_\nu^a)}{\delta \omega^b})~
e^{i\int d^4x [-\frac{1}{4}{F^a}_{\mu \nu}^2[Q] -\frac{1}{2 \alpha} (\frac{\eta^\mu \eta^\nu}{\eta^2}\partial_\mu Q_\nu^a)^2+{\bar \psi} [i\gamma^\mu \partial_\mu -m +gT^a\gamma^\mu Q^a_\mu] \psi  ]}.
\label{cfq5ng}
\eea

The generating functional in the background field method of QCD with general non-covariant gauge fixing is given by \cite{noncov,noncov1}
\bea
&& Z[A,J,\eta,{\bar \eta}]=\int [dQ] [d{\bar \psi}] [d \psi ] ~{\rm det}(\frac{\delta {\cal G}^a(Q)}{\delta \omega^b}) \nonumber \\
&& \times e^{i\int d^4x [-\frac{1}{4}{F^a}_{\mu \nu}^2[A+Q] -\frac{1}{2 \alpha}
({\cal G}^a(Q))^2+{\bar \psi} [i\gamma^\mu \partial_\mu -m +gT^a\gamma^\mu (A+Q)^a_\mu] \psi + J \cdot Q +{\bar \eta} \psi +{\bar \psi} \eta  ]}
\label{azaqcdng}
\eea
where
\bea
{\cal G}^a(Q) =\frac{\eta^\mu \eta^\nu}{\eta^2} ~(\partial_\mu Q_\nu^a + gf^{abc} A_\mu^b Q_\nu^c)=\frac{\eta^\mu \eta^\nu}{\eta^2}~D_\mu[A]Q_\nu^a.
\label{gang}
\eea

The non-perturbative correlation function of the type $<0|{\bar \psi}(x_2) \psi(x_1)|0>_A$
in the background field method of QCD in general non-covariant gauge is given by
\bea
&&<0|{\bar \psi}(x_2) \psi(x_1)|0>_A=\int [dQ] [d{\bar \psi}] [d \psi ] ~{\bar \psi}(x_2) \psi(x_1)\nonumber \\
&& \times {\rm det}(\frac{\delta {\cal G}^a(Q)}{\delta \omega^b}) e^{i\int d^4x [-\frac{1}{4}{F^a}_{\mu \nu}^2[A+Q] -\frac{1}{2 \alpha}
({\cal G}^a(Q))^2+{\bar \psi} [i\gamma^\mu \partial_\mu -m +gT^a\gamma^\mu (A+Q)^a_\mu] \psi   ]}.
\label{cfqcdng}
\eea

By changing the integration variable $Q \rightarrow Q-A$ in the right hand side of eq. (\ref{cfqcdng}) we find
\bea
&& <0|{\bar \psi}(x_2) \psi(x_1)|0>_A= \int [dQ] [d{\bar \psi}] [d \psi ]  ~~{\bar \psi}(x_2)\psi(x_1) \nonumber \\
&& \times {\rm det}(\frac{\delta {\cal G}_f^a(Q)}{\delta \omega^b})~~
e^{i\int d^4x [-\frac{1}{4}{F^a}_{\mu \nu}^2[Q] -\frac{1}{2 \alpha} ({\cal G}_f^a(Q))^2+{\bar \psi} [i\gamma^\mu \partial_\mu -m +gT^a\gamma^\mu Q^a_\mu] \psi  ]}
\label{cfqcd1ng}
\eea
where from eq. (\ref{gang}) we find
\bea
&&{\cal G}_f^a(Q) =\frac{\eta^\mu \eta^\nu}{\eta^2}~(\partial_\mu Q_\nu^a + gf^{abc} A_\mu^b Q_\nu^c - \partial_\mu A_\nu^a)-\frac{1}{\eta^2}~gf^{abc} (\eta \cdot A^b) (\eta \cdot A^c) \nonumber \\
&& =\frac{\eta^\mu \eta^\nu}{\eta^2}~(D_\mu[A] Q_\nu^a) - \frac{\eta^\mu \eta_\nu}{\eta^2}~\partial_\mu A_\nu^a.
\label{gfang}
\eea

Changing the integration variable from unprimed variable to primed variable we find from eq. (\ref{cfqcd1ng})
\bea
&& <0|{\bar \psi}(x_2) \psi(x_1)|0>_A= \int [dQ'] [d{\bar \psi}'] [d \psi' ]~~{\bar \psi}'(x_2)\psi'(x_1) \nonumber \\
&& \times {\rm det}(\frac{\delta {\cal G}_f^a(Q')}{\delta \omega^b})~~
e^{i\int d^4x [-\frac{1}{4}{F^a}_{\mu \nu}^2[Q'] -\frac{1}{2 \alpha} ({\cal G}_f^a(Q'))^2+{\bar \psi}' [i\gamma^\mu \partial_\mu -m +gT^a\gamma^\mu Q'^a_\mu] \psi'  ]}.
\label{cfqcd1vbng}
\eea
This is because a change of integration variable from unprimed variable to primed variable does not change the value of the
integration.

Using eqs. (\ref{dqa}), (\ref{psa}) and (\ref{tej}) in eq. (\ref{cfqcd1vbng}) we find
\bea
&& <0|{\bar \psi}(x_2) \psi(x_1)|0>_A=~\int [dQ] [d{\bar \psi}] [d \psi ] ~~{\bar \psi}(x_2)[{\cal P}e^{-igT^a \int_{x_1}^{x_2} dx^\mu A_\mu^a(x)}]\psi(x_1)\nonumber \\
 && \times {\rm det}(\frac{\delta {\cal G}_f^a(Q')}{\delta \omega^b})
~e^{i\int d^4x [-\frac{1}{4}{F^a}_{\mu \nu}^2[Q] -\frac{1}{2 \alpha} ({\cal G}_f^a(Q'))^2+{\bar \psi} [i\gamma^\mu \partial_\mu -m +gT^a\gamma^\mu Q^a_\mu] \psi ]}.
\label{cfqcd1cng}
\eea

From eq. (\ref{gfang}) we find
\bea
{\cal G}_f^a(Q') =
\frac{\eta^\mu \eta^\nu}{\eta^2}[\partial_\mu Q'^a_\nu + gf^{abc} A_\mu^b Q'^c_\nu] - \frac{\eta^\mu \eta^\nu}{\eta^2}~\partial_\mu A^a_\nu.
\label{gfapng}
\eea
By using eqs. (\ref{teq}) and (\ref{pg4}) in eq. (\ref{gfapng}) we find
\bea
&&{\cal G}_f^a(Q') =\frac{\eta^\mu \eta^\nu}{\eta^2}[\partial_\mu [[e^{gM(x)}]_{ab}Q_\nu^b(x) ~+ ~[\frac{e^{gM(x)}-1}{gM(x)}]_{ab}~[\partial_\nu \omega^b(x)]]\nonumber \\
&&+ gf^{abc} [\partial_\mu \omega^e(x)\left[\frac{e^{gM(x)}-1}{gM(x)}\right]_{be}] [[e^{gM(x)}]_{cd}Q_\nu^d(x) ~+ ~[\frac{e^{gM(x)}-1}{gM(x)}]_{cd}~[\partial_\nu \omega^d(x)]]]\nonumber \\
&&-\frac{\eta^\mu \eta^\nu}{\eta^2} \partial_\mu [\partial_\nu \omega^b(x)\left[\frac{e^{gM(x)}-1}{gM(x)}\right]_{ab}]
\label{gfapang}
\eea
which gives
\bea
&&{\cal G}_f^a(Q') =\frac{\eta^\mu \eta^\nu}{\eta^2}[\partial_\mu [[e^{gM(x)}]_{ab}Q_\nu^b(x)]\nonumber \\
&&+ gf^{abc} [\partial_\mu \omega^e(x)\left[\frac{e^{gM(x)}-1}{gM(x)}\right]_{be}] [[e^{gM(x)}]_{cd}Q_\nu^d(x) ~+ ~[\frac{e^{gM(x)}-1}{gM(x)}]_{cd}~[\partial_\nu \omega^d(x)]]].
\label{gfapbng}
\eea
From eq. (\ref{gfapbng}) we find
\bea
&&{\cal G}_f^a(Q') =\frac{\eta^\mu \eta^\nu}{\eta^2}[\partial_\mu [[e^{gM(x)}]_{ab}Q_\nu^b(x)]\nonumber \\
&&+ gf^{abc} [[\partial_\mu \omega^e(x)]\left[\frac{e^{gM(x)}-1}{gM(x)}\right]_{be}] [[e^{gM(x)}]_{cd}Q_\nu^d(x)]]
\label{gfapcng}
\eea
which gives
\bea
&&{\cal G}_f^a(Q') =\frac{\eta^\mu \eta^\nu}{\eta^2} [[e^{gM(x)}]_{ab}\partial_\mu Q_\nu^b(x)+Q_\mu^b(x)\partial_\nu [[e^{gM(x)}]_{ab}]\nonumber \\
&&+  [[\partial_\mu \omega^e(x)]\left[\frac{e^{gM(x)}-1}{gM(x)}\right]_{be}]gf^{abc} [[e^{gM(x)}]_{cd}Q_\nu^d(x)]].
\label{gfapdng}
\eea
Using eq. (\ref{pg4k}) in (\ref{gfapdng}) we find
\bea
&&{\cal G}_f^a(Q') = \frac{\eta^\mu \eta^\nu}{\eta^2}[e^{gM(x)}]_{ab}\partial_\mu Q_\nu^b(x)
\label{gfapeng}
\eea
which gives
\bea
({\cal G}_f^a(Q'))^2 = (\frac{\eta^\mu \eta^\nu}{\eta^2}\partial_\mu Q_\nu^a(x))^2.
\label{gfapfng}
\eea
From eqs. (\ref{detr}) and (\ref{gfapeng}) we find
\bea
&&{\rm det} [\frac{\delta {\cal G}_f^a(Q')}{\delta \omega^b}] ={\rm det}
[\frac{\eta^\mu \eta^\nu}{\eta^2}\frac{ \delta [[e^{gM(x)}]_{ac}\partial_\mu Q_\nu^c(x)]}{\delta \omega^b}]={\rm det}[\frac{\eta^\mu \eta^\nu}{\eta^2}
[e^{gM(x)}]_{ac}\frac{ \delta (\partial_\mu Q_\nu^c(x))}{\delta \omega^b}]\nonumber \\
&&=\left[{\rm det}[
[e^{gM(x)}]_{ac}]\right]~\left[{\rm det}[\frac{\eta^\mu \eta^\nu}{\eta^2}\frac{ \delta (\partial_\mu Q_\nu^c(x))}{\delta \omega^b}]\right]={\rm exp}[{\rm Tr}({\rm ln}[e^{gM(x)}])]~{\rm det}[\frac{\eta^\mu \eta^\nu}{\eta^2}\frac{ \delta (\partial_\mu Q_\nu^a(x))}{\delta \omega^b}]\nonumber \\
&&={\rm det}[\frac{\eta^\mu \eta^\nu}{\eta^2}\frac{ \delta (\partial_\mu Q_\nu^a(x))}{\delta \omega^b}].
\label{gqp4ang}
\eea
Using eqs. (\ref{gfapfng}) and (\ref{gqp4ang}) in eq. (\ref{cfqcd1cng}) we find
\bea
&&<0|{\bar \psi}(x_2) \psi(x_1)|0>_A=\int [dQ] [d{\bar \psi}] [d \psi ] ~~{\bar \psi}(x_2)[{\cal P}e^{-igT^a \int_{x_1}^{x_2} dx^\mu A_\mu^a(x)}]\psi(x_1)\nonumber \\
 && \times
{\rm det}(\frac{\eta^\mu \eta^\nu}{\eta^2}\frac{\delta (\partial_\mu Q_\nu^a)}{\delta \omega^b})~
e^{i\int d^4x [-\frac{1}{4}{F^a}_{\mu \nu}^2[Q] -\frac{1}{2 \alpha}(\frac{\eta^\mu \eta^\nu}{\eta^2}\partial_\mu Q_\nu^a)^2+{\bar \psi} [i\gamma^\mu \partial_\mu -m +gT^a\gamma^\mu Q^a_\mu] \psi  ]}.
\label{cfq5p1ng}
\eea
Using the similar technique as above we find
\bea
&&<0|~{\bar \psi}(x_2)[{\cal P}e^{igT^a \int_{x_1}^{x_2} dx^\mu A_\mu^a(x)}]\psi(x_1)|0>_A
 =\int [dQ] [d{\bar \psi}] [d \psi ] ~{\bar \psi}(x_2) \psi(x_1)\nonumber \\
 && \times {\rm det}(\frac{\eta^\mu \eta^\nu}{\eta^2}\frac{\delta (\partial_\mu Q_\nu^a)}{\delta \omega^b})~
e^{i\int d^4x [-\frac{1}{4}{F^a}_{\mu \nu}^2[Q] -\frac{1}{2 \alpha}(\frac{\eta^\mu \eta^\nu}{\eta^2}\partial_\mu Q_\nu^a)^2+{\bar \psi} [i\gamma^\mu \partial_\mu -m +gT^a\gamma^\mu Q^a_\mu] \psi  ]}
\label{cfq5p2ng}
\eea
in general non-covariant gauge
in the presence of SU(3) pure gauge background field $A^{\mu a}(x)$ as given by eq. (\ref{gtqcd}).

From eqs. (\ref{lkjn}), (\ref{cfq5ng}) and (\ref{cfq5p2ng}) we find
\bea
&&<0|{\bar \psi}(x_2) \psi(x_1)|0> =<0|{\bar \psi}(x_2) [{\cal P}{\rm exp}[-ig\int_0^{\infty} d\lambda l\cdot { A}^c(x_2+l\lambda)T^{c}]]\nonumber \\
&&\times {\cal P}{\rm exp}[ig\int_0^{\infty} d\lambda l\cdot { A}^b(x_1+l\lambda)T^{b}]\psi_(x_1)|0>_A
\label{finalzng}
\eea
which proves factorization of soft-collinear divergences at all order in coupling constant in QCD when light-like Wilson line is
used in general non-covariant gauges.

\subsection{ Proof of Factorization Theorem in QCD With Light-Like Wilson Line in General Coulomb Gauge }

In QCD the generating functional with general Coulomb gauge fixing is given by \cite{noncov,noncov1}
\bea
&& Z[J,\eta,{\bar \eta}]=\int [dQ] [d{\bar \psi}] [d \psi ] ~{\rm det}(\frac{\delta ([g^{\mu \nu}-\frac{n^\mu n^\nu}{n^2}]\partial_\mu Q_\nu^a)}{\delta \omega^b}) \nonumber \\
&& \times e^{i\int d^4x [-\frac{1}{4}{F^a}_{\mu \nu}^2[Q] -\frac{1}{2 \alpha} ([g^{\mu \nu}-\frac{n^\mu n^\nu}{n^2}]\partial_\mu Q_\nu^a)^2+{\bar \psi} [i\gamma^\mu \partial_\mu -m +gT^a\gamma^\mu Q^a_\mu]  \psi + J \cdot Q +{\bar \psi} \eta +{\bar \eta} \psi ]}
\label{zfqc}
\eea
where
\bea
n^\mu =(1,0,0,0).
\eea
The non-perturbative correlation function of the type $<0|{\bar \psi}(x_2) \psi(x_1)|0>$ in QCD in general Coulomb gauge
is given by
\bea
&&<0|{\bar \psi}(x_2) \psi(x_1)|0>=\int [dQ] [d{\bar \psi}] [d \psi ] ~{\bar \psi}(x_2) \psi(x_1)\nonumber \\
&& \times ~{\rm det}(\frac{\delta ([g^{\mu \nu}-\frac{n^\mu n^\nu}{n^2}]\partial_\mu Q_\nu^a)}{\delta \omega^b})~
e^{i\int d^4x [-\frac{1}{4}{F^a}_{\mu \nu}^2[Q] -\frac{1}{2 \alpha} ([g^{\mu \nu}-\frac{n^\mu n^\nu}{n^2}]\partial_\mu Q_\nu^a)^2+{\bar \psi} [i\gamma^\mu \partial_\mu -m +gT^a\gamma^\mu Q^a_\mu] \psi  ]}.\nonumber \\
\label{cfq5c}
\eea
The generating functional in the background field method of QCD with general Coulomb gauge fixing is given by \cite{noncov,noncov1}
\bea
&& Z[A,J,\eta,{\bar \eta}]=\int [dQ] [d{\bar \psi}] [d \psi ] ~{\rm det}(\frac{\delta {\cal G}^a(Q)}{\delta \omega^b}) \nonumber \\
&& \times e^{i\int d^4x [-\frac{1}{4}{F^a}_{\mu \nu}^2[A+Q] -\frac{1}{2 \alpha}
({\cal G}^a(Q))^2+{\bar \psi} [i\gamma^\mu \partial_\mu -m +gT^a\gamma^\mu (A+Q)^a_\mu] \psi + J \cdot Q +{\bar \eta} \psi +{\bar \psi} \eta  ]}
\label{azaqcdc}
\eea
where
\bea
{\cal G}^a(Q) = [g^{\mu \nu}-\frac{n^\mu n^\nu}{n^2}](\partial_\mu Q_\nu^a + gf^{abc} A_\mu^b Q_\nu^c)=[g^{\mu \nu}-\frac{n^\mu n^\nu}{n^2}]D_\mu[A]Q_\nu^a.
\label{gac}
\eea
The non-perturbative correlation function of the type $<0|{\bar \psi}(x_2) \psi(x_1)|0>_A$
in the background field method of QCD in general Coulomb gauge is given by
\bea
&&<0|{\bar \psi}(x_2) \psi(x_1)|0>_A=\int [dQ] [d{\bar \psi}] [d \psi ] ~ {\bar \psi}(x_2) \psi(x_1)\nonumber \\
&& \times {\rm det}(\frac{\delta {\cal G}^a(Q)}{\delta \omega^b}) e^{i\int d^4x [-\frac{1}{4}{F^a}_{\mu \nu}^2[A+Q] -\frac{1}{2 \alpha}
({\cal G}^a(Q))^2+{\bar \psi} [i\gamma^\mu \partial_\mu -m +gT^a\gamma^\mu (A+Q)^a_\mu] \psi   ]}.
\label{cfqcdc}
\eea
Hence by replacing $\frac{\eta^\mu \eta^\nu}{\eta^2} \rightarrow [g^{\mu \nu}-\frac{n^\mu n^\nu}{n^2}]$ everywhere in the derivations
in the previous sub-section we find
\bea
&&<0|{\bar \psi}(x_2) \psi(x_1)|0> =<0|{\bar \psi}(x_2) [{\cal P}{\rm exp}[-ig\int_0^{\infty} d\lambda l\cdot { A}^c(x_2+l\lambda)T^{c}]]\nonumber \\
&&\times {\cal P}{\rm exp}[ig\int_0^{\infty} d\lambda l\cdot { A}^b(x_1+l\lambda)T^{b}]\psi_(x_1)|0>_A
\label{finalzc}
\eea
which proves factorization of soft-collinear divergences at all order in coupling constant in QCD when light-like Wilson line is
used in general Coulomb gauge.

\section{ Violation of Factorization Theorem of Soft-Collinear Divergences in QCD With non-light-like Wilson line }

As mentioned earlier, in some of the studies, the non-light-like Wilson line is used in the definition of the transverse
momentum dependent parton distribution function (TMD PDF) at high energy colliders \cite{collins,collinstmd}.
These studies involve diagrammatic calculation at one loop level using perturbative QCD.
However, the (transverse momentum dependent) parton distribution function and fragmentation function
are non-perturbative quantities in QCD. Hence if one uses the perturbation theory to study
their properties \cite{collins,collinstmd}, then one may end up finding wrong results.
In general, the non-perturbative phenomena may be impossible to understand by
perturbation theory, regardless of how many orders of perturbation theory one uses. Take for example,
the non-perturbative function
\bea
f(x) =e^{-\frac{1}{x^{4}}}.
\eea
The Taylor series at $x = 0$ for this function $f(x)$ is exactly zero to all orders in perturbation theory,
but the function is non-zero if $x \neq 0$.

Hence the diagrammatic method using perturbative QCD may not be always sufficient to prove factorization of
soft-collinear divergences of the (transverse momentum dependent) parton distribution function and fragmentation function
at high energy colliders,  regardless of how many orders of perturbation theory one uses,
because the (transverse momentum dependent) parton distribution function and fragmentation function are non-perturbative quantities in QCD.

On the other hand the path integral method of QCD can be used to study non-perturbative QCD.
The path integral method of QCD is valid at all order in coupling constant. Hence we find
that the path integral method of QCD is very useful to prove factorization of
soft-collinear divergences of non-perturbative quantities in QCD such as the
(transverse momentum dependent) parton distribution function and fragmentation function at all order in coupling constant
at high energy colliders.

In this paper we have proved in section VII, by using path integral method of QCD, that the factorization theorem in QCD is valid
at all order in coupling constant when light-like Wilson line is used. In this section we will prove, by using path integral
method of QCD, that the factorization theorem in QCD is not valid
at all order in coupling constant when non-light-like Wilson line is used.
This implies that the factorization theorem is violated in all the previous studies \cite{collins,collinstmd}
which used the non-light-like Wilson line in the definition of  the (transverse momentum dependent)
parton distribution function and fragmentation function at high energy colliders.
This is in conformation with the finding in \cite{nayaksterman,nayaksterman1} which proved factorization
theorem in NRQCD heavy quarkonium production in case of light-like Wilson line
\cite{nayaksterman} and the violation of factorization theorem in NRQCD heavy quarkonium production in case of
non-light-like Wilson line \cite{nayaksterman1}. In case of massive Wilson line in QCD the color transfer occurs
and the factorization breaks down \cite{nayaksterman1}. The simple physics reason behind this is the following.
The light-like Wilson line produces pure gauge potential which gives $F_{\mu \nu}^a(x) = 0$ which can be gauged away
in the sense of factorization because pure gauge corresponds to unphysical longitudinal polarization
and hence factorization theorem works (see sections IV and VII for details).
The non-light-like Wilson line does not produce pure gauge potential ($F_{\mu \nu}^a(x) \neq 0$),
the form of which can be arbitrary which can not be gauged away
in the sense of factorization because non-pure gauge does not correspond to unphysical longitudinal
polarization and hence the non-light-like Wilson line spoils the factorization (see sections V and
this section for details).

\subsection{ Violation of Factorization Theorem in QCD with Non-Light-Like Wilson Line In Covariant Gauge }

Note that eq. (\ref{cfqcd1}) is valid for any arbitrary background field $A^{\mu a}(x)$. Hence we
start from eq. (\ref{cfqcd1}) by following exactly the same procedure that was followed for light-like
Wilson line case. By changing the integration variable from unprimed variable to primed variable we find from eq. (\ref{cfqcd1})
\bea
&& <0|{\bar \psi}(x_2)\psi(x_1)|0>_A= \int [dQ'] [d{\bar \psi}'] [d \psi' ]~{\bar \psi}'(x_2)\psi'(x_1) \nonumber \\
&& \times {\rm det}(\frac{\delta G_f^a(Q')}{\delta \omega^b})~~
e^{i\int d^4x [-\frac{1}{4}{F^a}_{\mu \nu}^2[Q'] -\frac{1}{2 \alpha} (G_f^a(Q'))^2+{\bar \psi}' [i\gamma^\mu \partial_\mu -m +gT^a\gamma^\mu Q'^a_\mu] \psi'  ]}.
\label{cfqcd1vbn}
\eea
This is because a change of integration variable from unprimed variable to primed variable does not change the value of the
integration.

Now let us define the primed variables $Q'$ and $\psi'$ for the non-light-like Wilson line case  in analogous to the light-like Wilson line case.
First of all in analogous to eq. (\ref{ppgn}) for the light-like Wilson line
case let us define the primed variable $\psi'$ for the non-light-like Wilson line case
\bea
\psi'(x) = {\cal P}e^{-ig \int_0^{\infty} d\lambda v\cdot { A}^a(x+v\lambda)T^a }\psi(x)
\label{ppgnn}
\eea
where $A^{\mu a}(x)$ is not the SU(3) pure gauge and $v^\mu$ is the non-light-like four velocity.
Similarly in analogous to eq. (\ref{ftgrm}) for the light-like Wilson line case let us define the primed variable $Q'$
for the non-light-like Wilson line case
\bea
&&T^aQ'^a_\mu(x) = U(x)T^aQ^a_\mu(x) U^{-1}(x)+\frac{1}{ig}[\partial_\mu U(x)] U^{-1}(x)
\label{ftgrmn}
\eea
where
\bea
U(x)={\cal P}e^{-ig \int_0^{\infty} d\lambda v\cdot { A}^a(x+v\lambda)T^a }
\label{ftgrmun}
\eea
where $A^{\mu a}(x)$ is not the SU(3) pure gauge and $v^\mu$ is the non-light-like four velocity.
Under this finite transformation, using eq. (\ref{ftgrmn}), we find
\bea
&& [dQ'] =[dQ] ~{\rm det} [\frac{\partial {Q'}^a}{\partial Q^b}] = [dQ]
\label{dqan}
\eea
which is similar to eq. (\ref{dqa}) for the light-like Wilson line case.
From eqs. (\ref{ftgrmn}), (\ref{ftgrmun}) and (\ref{ppgnn}) we find that
\bea
&&[d{\bar \psi}'] [d \psi' ]=[d{\bar \psi}] [d \psi ],~~~~~~{\bar \psi}' [i\gamma^\mu \partial_\mu -m +gT^a\gamma^\mu Q'^a_\mu] \psi'={\bar \psi} [i\gamma^\mu \partial_\mu -m +gT^a\gamma^\mu Q^a_\mu]\psi,\nonumber \\
&&{F^a}_{\mu \nu}^2[Q']={F^a}_{\mu \nu}^2[Q]
\label{psan}
\eea
for the non-light-like Wilson line case which is similar to eq. (\ref{psa}) for the light-like Wilson line case.

Using eqs. (\ref{dqan}) and (\ref{psan}) in eq. (\ref{cfqcd1vbn}) we find
\bea
&& <0|{\bar \psi}(x_2)\psi(x_1)|0>_A=~\int [dQ] [d{\bar \psi}] [d \psi ] {\bar \psi}(x_2)[[{\cal P}{\rm exp}[ig\int_0^{\infty} d\lambda v\cdot { A}^c(x_2+v\lambda)T^{c}]]\nonumber \\
 && \times {\cal P}{\rm exp}[-ig\int_0^{\infty} d\lambda v\cdot { A}^b(x_1+v\lambda)T^{b}]] \psi(x_1)~\nonumber \\
 && \times {\rm det}(\frac{\delta G_f^a(Q')}{\delta \omega^b})
~e^{i\int d^4x [-\frac{1}{4}{F^a}_{\mu \nu}^2[Q] -\frac{1}{2 \alpha} (G_f^a(Q'))^2+{\bar \psi} [i\gamma^\mu \partial_\mu -m +gT^a\gamma^\mu Q^a_\mu] \psi ]}.
\label{cfqcd1cn}
\eea
From eq. (\ref{gfa}) we find
\bea
G_f^a(Q') =\partial_\mu Q^{' \mu a} + gf^{abc} A_\mu^b Q^{' \mu c} - \partial_\mu A^{\mu a}.
\label{gfapn}
\eea
Now from eqs. (\ref{ftgrmn}) and (\ref{ftgrmun}) we find that for the non-light-like Wilson line case
\bea
{Q'}_\mu^a(x) -A_\mu^a(x)\neq [e^{gN(x)}]_{ab}Q_\mu^b(x),~~~~~~~~~~~~~~~~~~~~~~~N_{ab}(x)=f^{abc}h^c(x)
\label{tejn}
\eea
when $A^{\mu a}(x)$ is not the SU(3) pure gauge where $h^a(x)$ is any function.
Hence we find that eq. (\ref{tejn}) for the non-light-like
Wilson line case differs from the corresponding eq. (\ref{tej}) for the light-like Wilson line case. From
eqs. (\ref{gfapn}), (\ref{ftgrmn}), (\ref{ftgrmun}) and (\ref{tejn}) we find
\bea
&&G_f^a(Q') \neq [e^{gS(x)}]_{ab}\partial^\mu Q_\mu^b(x),~~~~~~~~~~~~~~~~~~~~~~~S_{ab}(x)=f^{abc}w^c(x)
\label{gfapnen}
\eea
where $w^a(x)$ is any function which gives
\bea
(G_f^a(Q'))^2 \neq (\partial_\mu Q^{\mu a}(x))^2
\label{gfapfn}
\eea
for the non-light-like Wilson line case which differs from the corresponding eq. (\ref{gfapf}) for
the light-like Wilson line case. From eq. (\ref{gfapnen})  we find
\bea
&&{\rm det} [\frac{\delta G_f^a(Q')}{\delta \omega^b}]  \neq {\rm det}[\frac{ \delta (\partial_\mu Q^{\mu a}(x))}{\delta \omega^b}]
\label{gqp4an}
\eea
for the non-light-like Wilson line case which differs from the corresponding eq. (\ref{gqp4a}) for the light-like Wilson line case.

Hence we find from eqs. (\ref{cfqcd1cn}), (\ref{gfapfn}) and (\ref{gqp4an}) that
\bea
&& <0|{\bar \psi}(x_2)\psi(x_1)|0>_A\neq \int [dQ] [d{\bar \psi}] [d \psi ] {\bar \psi}(x_2)[[{\cal P}{\rm exp}[ig\int_0^{\infty} d\lambda v\cdot { A}^c(x_2+v\lambda)T^{c}]]\nonumber \\
 && \times {\cal P}{\rm exp}[-ig\int_0^{\infty} d\lambda v\cdot { A}^b(x_1+v\lambda)T^{b}]]\psi(x_1)~\nonumber \\
 && \times {\rm det}(\frac{\delta (\partial_\mu Q^{\mu a})}{\delta \omega^b})~
 e^{i\int d^4x [-\frac{1}{4}{F^a}_{\mu \nu}^2[Q] -\frac{1}{2 \alpha}(\partial_\mu Q^{\mu a})^2+{\bar \psi} [i\gamma^\mu \partial_\mu -m +gT^a\gamma^\mu Q^a_\mu] \psi  ]}
\label{cfqcd1cnn}
\eea
which gives
\bea
&& <0|{\bar \psi}(x_2)[[{\cal P}{\rm exp}[-ig\int_0^{\infty} d\lambda v\cdot { A}^c(x_2+v\lambda)T^{c}]]\times {\cal P}{\rm exp}[ig\int_0^{\infty} d\lambda v\cdot { A}^b(x_1+v\lambda)T^{b}]]\psi(x_1)|0>_A~\nonumber \\
 && \neq \int [dQ] [d{\bar \psi}] [d \psi ] {\bar \psi}(x_2)\psi(x_1)~{\rm det}(\frac{\delta (\partial_\mu Q^{\mu a})}{\delta \omega^b})~
 e^{i\int d^4x [-\frac{1}{4}{F^a}_{\mu \nu}^2[Q] -\frac{1}{2 \alpha}(\partial_\mu Q^{\mu a})^2+{\bar \psi} [i\gamma^\mu \partial_\mu -m +gT^a\gamma^\mu Q^a_\mu] \psi  ]}~\nonumber \\
\label{cfqcd1cno}
\eea
for the non-light-like Wilson line case which differs from the corresponding eq. (\ref{cfq5p2}) for the light-like Wilson line case.

Hence from eqs. (\ref{corq}) and (\ref{cfqcd1cno}) we find
\bea
&&<0|{\bar \psi}(x_2)\psi(x_1)|0> \nonumber \\
 && \neq <0|{\bar \psi}(x_2)[[{\cal P}{\rm exp}[-ig\int_0^{\infty} d\lambda v\cdot { A}^c(x_2+v\lambda)T^{c}]]\times {\cal P}{\rm exp}[ig\int_0^{\infty} d\lambda v\cdot { A}^b(x_1+v\lambda)T^{b}]]\psi(x_1)|0>_A \nonumber \\
\label{finalyn}
\eea
which proves that the factorization theorem is violated at all order in coupling constant when the non-light-like Wilson line is used in QCD
in covariant gauge where $v^\mu$ is the non-light-like four velocity.

\subsection{Violation of Factorization Theorem in QCD with Non-Light-Like Wilson Line In General Axial Gauge }

The non-perturbative correlation function of the type $<0|{\bar \psi}(x_1) \psi(x_2)|0>_A$
in the background field method of QCD in general axial gauge is given by eq. (\ref{cfqcd1ag})
which is valid for any arbitrary background field $A^{\mu a}(x)$.
Changing the integration variable from unprimed variable to primed variable we find from eq. (\ref{cfqcd1ag})
\bea
&& <0|{\bar \psi}(x_1) \psi(x_2)|0>_A= \int [dQ'] [d{\bar \psi}'] [d \psi' ]~{\bar \psi}'(x_1)\psi'(x_2) \nonumber \\
&& \times
e^{i\int d^4x [-\frac{1}{4}{F^a}_{\mu \nu}^2[Q'] -\frac{1}{2 \alpha} (\eta^\mu (Q'-A)_\mu^a)^2+{\bar \psi}' [i\gamma^\mu \partial_\mu -m +gT^a\gamma^\mu Q'^a_\mu] \psi'  ]}.
\label{cfqcd1vbagv}
\eea
This is because a change of integration variable from unprimed variable to primed variable does not change the value of the
integration.

Now from eqs. (\ref{psan}), (\ref{tejn}), (\ref{cfqcd1vbagv}) and (\ref{cfq5ag}) one finds in the general axial gauge
\bea
&&<0|{\bar \psi}(x_2)\psi(x_1)|0> \nonumber \\
 && \neq <0|{\bar \psi}(x_2)[[{\cal P}{\rm exp}[-ig\int_0^{\infty} d\lambda v\cdot { A}^c(x_2+v\lambda)T^{c}]]\times {\cal P}{\rm exp}[ig\int_0^{\infty} d\lambda v\cdot { A}^b(x_1+v\lambda)T^{b}]]\psi(x_1)|0>_A \nonumber \\
\label{finalynga}
\eea
which proves that the factorization theorem is violated at all order in coupling constant when the non-light-like Wilson line is
used in QCD in the general axial gauge where $v^\mu$ is the non-light-like four velocity.

\subsection{ Violation of Factorization Theorem in QCD with Non-Light-Like Wilson Line In Light Cone Gauge }

Since eq. (\ref{finalynga}) is valid for general axial gauges [see eq. (\ref{axialg})]
it is also valid for light cone gauge ($\eta^2=0$). Hence from eq. (\ref{finalynga}) we find that in the light cone gauge
\bea
&&<0|{\bar \psi}(x_2)\psi(x_1)|0> \nonumber \\
 && \neq <0|{\bar \psi}(x_2)[[{\cal P}{\rm exp}[-ig\int_0^{\infty} d\lambda v\cdot { A}^c(x_2+v\lambda)T^{c}]]\times {\cal P}{\rm exp}[ig\int_0^{\infty} d\lambda v\cdot { A}^b(x_1+v\lambda)T^{b}]]\psi(x_1)|0>_A \nonumber \\
\label{finalyngalc}
\eea
which proves that the factorization theorem is violated at all order in coupling constant when the non-light-like Wilson line is
used in QCD in the light cone gauge where $v^\mu$ is the non-light-like four velocity.

\subsection{ Violation of Factorization Theorem in QCD with Non-Light-Like Wilson Line In General Non-Covariant Gauge }

The non-perturbative correlation function of the type $<0|{\bar \psi}(x_1) \psi(x_2)|0>_A$
in the background field method of QCD in general non-covariant gauges is given by eq. (\ref{cfqcd1ng})
which is valid for any arbitrary background field $A^{\mu a}(x)$.
Changing the integration variable from unprimed variable to primed variable we find from eq. (\ref{cfqcd1ng})
\bea
&& <0|{\bar \psi}(x_1) \psi(x_2)|0>_A= \int [dQ'] [d{\bar \psi}'] [d \psi' ]~~{\bar \psi}'(x_1)\psi'(x_2) \nonumber \\
&& \times {\rm det}(\frac{\delta {\cal G}_f^a(Q')}{\delta \omega^b})~~
e^{i\int d^4x [-\frac{1}{4}{F^a}_{\mu \nu}^2[Q'] -\frac{1}{2 \alpha} ({\cal G}_f^a(Q'))^2+{\bar \psi}' [i\gamma^\mu \partial_\mu -m +gT^a\gamma^\mu Q'^a_\mu] \psi'  ]}.
\label{cfqcd1vbngvx}
\eea
This is because a change of integration variable from unprimed variable to primed variable does not change the value of the
integration. From eqs. (\ref{gfang}), (\ref{ftgrmn}), (\ref{ftgrmun}) and (\ref{tejn}) we find
\bea
&&{\cal G}_f^a(Q') \neq [e^{gS(x)}]_{ab}\frac{\eta^\mu \eta^\nu}{\eta^2}~(\partial_\mu  Q_\nu^b(x),~~~~~~~~~~~~~~~~~~~~~~~S_{ab}(x)=f^{abc}w^c(x)
\label{gfapnenv}
\eea
where $w^a(x)$ is any function which gives
\bea
({\cal G}_f^a(Q'))^2 \neq (\frac{\eta^\mu \eta^\nu}{\eta^2}~\partial_\mu  Q_\nu^a(x))^2
\label{gfapfnv}
\eea
for the non-light-like Wilson line case which differs from the corresponding eq. (\ref{gfapfng}) for
the light-like Wilson line case. From eq. (\ref{gfapnenv})  we find
\bea
&&{\rm det} [\frac{\delta {\cal G}_f^a(Q')}{\delta \omega^b}]  \neq {\rm det}[\frac{ \delta (\frac{\eta^\mu \eta^\nu}{\eta^2}~\partial_\mu Q_\nu^a(x))}{\delta \omega^b}]
\label{gqp4anv}
\eea
for the non-light-like Wilson line case which differs from the corresponding eq. (\ref{gqp4ang}) for the light-like Wilson line case.

Now from eqs. (\ref{psan}), (\ref{tejn}), (\ref{gfapfnv}), (\ref{gqp4anv}),
(\ref{cfqcd1vbngvx}) and (\ref{cfq5ng}) we find that in the general non-covariant gauge
\bea
&&<0|{\bar \psi}(x_2)\psi(x_1)|0> \nonumber \\
 && \neq <0|{\bar \psi}(x_2)[[{\cal P}{\rm exp}[-ig\int_0^{\infty} d\lambda v\cdot { A}^c(x_2+v\lambda)T^{c}]]\times {\cal P}{\rm exp}[ig\int_0^{\infty} d\lambda v\cdot { A}^b(x_1+v\lambda)T^{b}]]\psi(x_1)|0>_A \nonumber \\
\label{finalyngavnc}
\eea
which proves that the factorization theorem is violated at all order in coupling constant when the non-light-like Wilson line is
used in QCD in the general non-covariant gauge where $v^\mu$ is the non-light-like four velocity.

\subsection{ Violation of Factorization Theorem in QCD with Non-Light-Like Wilson Line In General Coulomb Gauge }

The non-perturbative correlation function of the type $<0|{\bar \psi}(x_1) \psi(x_2)|0>_A$
in the background field method of QCD in general Coulomb gauge is given by eq. (\ref{cfqcdc})
which is valid for any arbitrary background field $A^{\mu a}(x)$ where
\bea
{\cal G}^a(Q) = [g^{\mu \nu}-\frac{n^\mu n^\nu}{n^2}](\partial_\mu Q_\nu^a + gf^{abc} A_\mu^b Q_\nu^c)=[g^{\mu \nu}-\frac{n^\mu n^\nu}{n^2}]D_\mu[A]Q_\nu^a.
\eea
and
\bea
n^\mu =(1,0,0,0).
\eea
Hence by replacing $\frac{\eta^\mu \eta^\nu}{\eta^2} \rightarrow [g^{\mu \nu}-\frac{n^\mu n^\nu}{n^2}]$ everywhere in the derivations
in the previous sub-section we find that in the general Coulomb gauge
\bea
&&<0|{\bar \psi}(x_2)\psi(x_1)|0> \nonumber \\
 && \neq <0|{\bar \psi}(x_2)[[{\cal P}{\rm exp}[-ig\int_0^{\infty} d\lambda v\cdot { A}^c(x_2+v\lambda)T^{c}]]\times {\cal P}{\rm exp}[ig\int_0^{\infty} d\lambda v\cdot { A}^b(x_1+v\lambda)T^{b}]]\psi(x_1)|0>_A \nonumber \\
\label{finalyngavncvc}
\eea
which proves that the factorization theorem is violated at all order in coupling constant when the non-light-like Wilson line is
used in QCD in the general Coulomb gauge where $v^\mu$ is the non-light-like four velocity.

Hence we have proved in this paper, by using path integral
method of QCD, that the soft-collinear divergences are factorized at all order in coupling constant when light-like Wilson line is used
in the definition of the parton distribution function and fragmentation function at high energy colliders
but the soft-collinear divergences are not factorized when the non-light-like Wilson line is used
in the definition of the (transverse momentum dependent)
parton distribution function and fragmentation function at high energy colliders.
This is in conformation with the finding in \cite{nayaksterman,nayaksterman1} which used the diagrammatic
method of QCD to prove factorization theorem in NRQCD heavy quarkonium production in case of light-like Wilson line
\cite{nayaksterman} and violation of factorization theorem in NRQCD heavy quarkonium production in case of
non-light-like Wilson line \cite{nayaksterman1}. In case of massive Wilson line in QCD the color transfer occurs
and the factorization breaks down \cite{nayaksterman1}.

Hence we find that the factorization theorem is violated in all the previous studies \cite{collins,collinstmd}
which used the non-light-like Wilson line in the definition of the (transverse momentum dependent)
parton distribution function and fragmentation function at high energy colliders.

\section{Conclusions}

By using path integral formulation of QCD and QED we have proven that the factorization theorem is valid
for light-like Wilson line but is not valid for non-light-like Wilson line. This  conclusion is shown
to be consistent with Ward identity and Grammer-Yennie approximation. Hence we have found that
the factorization theorem is violated in all the previous studies which used the non-light-like
Wilson line in the definition of  the (transverse momentum dependent)
parton distribution function and fragmentation function at high energy colliders.

\end{document}